\documentstyle[multicol,amssymb,aps,epsfig]{revtex}

\begin{document}
\draft

\title{Trapping of single atoms with single photons in cavity QED}
\author{A.\ C. Doherty, T.\ W. Lynn, C.\ J. Hood, and H.\ J. Kimble}
\address{Norman Bridge Laboratory of Physics 12-33,\\
California Institute of Technology, Pasadena, California 91125}

\maketitle

\begin{abstract}
Two recent experiments have reported the trapping of individual atoms inside
optical resonators by the mechanical forces associated with single photons
[Hood et al., Science 287, 1447 (2000), Pinkse et al., Nature 404, 365
(2000)]. Here we analyze the trapping dynamics in these settings, focusing on
two points of interest. Firstly, we investigate the extent to which
light-induced forces in these experiments are distinct from their free-space
counterparts, and whether or not there are qualitatively different
effects of optical forces at the single-photon level within the
setting of cavity QED. Secondly, we
explore the quantitative features of the resulting 
atomic motion, and how these dynamics are mapped onto experimentally
observable variations of the
intracavity field. Towards these ends, we present results from extensive numerical
simulations of the relevant forces and their fluctuations, as well as a
detailed derivation of our numerical simulation method, based on the full
quantum-mechanical master equation. Not surprisingly, qualitatively distinct atomic dynamics
arise as the coupling and dissipative rates are varied. For the experiment of
Hood et al., we show that atomic motion is largely conservative and is
predominantly in radial orbits transverse to the cavity axis. A comparison
with the free-space theory demonstrates that the fluctuations of the dipole
force are suppressed by an order of magnitude. This effect is based upon the
Jaynes-Cummings eigenstates of the atom-cavity system and represents
distinct physics for optical forces at the single-photon level within the context of cavity QED. By
contrast, even in a regime of strong coupling in the experiment of Pinkse et
al., there are only small quantitative distinctions between the
potentials and heating rates in the free-space
theory and the quantum theory, so it is not clear that a description of this
experiment as a novel single-quantum trapping effect is necessary.  The
atomic motion is strongly diffusive, leading to an average localization time
comparable to the time for an atom to transit freely through the cavity, and
to a reduction in the ability to infer aspects of the atomic motion from the
intracavity photon number. 
\end{abstract}

\pacs{42.50.Vk,42.50.Ct,32.80.Pj }

\begin{multicols}{2}

\section{Introduction}

An exciting advance in recent years has been the increasing ability to
observe and manipulate the dynamical processes of individual quantum
systems. In this{\bf \ }endeavor, an important physical system has been a
single atom strongly coupled to the electromagnetic field of a high-$Q$
(optical or microwave) cavity within the setting of cavity quantum
electrodynamics (cavity QED). \cite{berman94,nobel} Here the coupling
frequency of one atom to a single mode of an optical resonator is denoted by
$g_{0}$ (i.e., $2g_{0}$ is the one-photon Rabi frequency), with the regime
of strong coupling defined by the requirement that $g_{0}\gg (\gamma ,\kappa
)$, where $\gamma $ is the atomic decay rate to modes other than the cavity
mode and $\kappa $ is the decay rate of the cavity mode itself. In this
circumstance, the number of photons required to saturate an intracavity atom
is $n_{0}\sim {\gamma ^{2}/g_{0}^{2}}\ll 1$ and the number of atoms
required to have an appreciable effect on the intracavity field is $%
N_{0}\sim \kappa \gamma /g_{0}^{2}\ll 1$ \cite{hjksweden}.

Although there have been numerous laboratory advances which demonstrate the
effect of strong coupling on the {\it internal} degrees of freedom of an
atomic dipole coupled to the quantized cavity field (i.e., $g_{0}\gg $ $%
\kappa ,$ $\gamma $), the consequences of strong coupling for the{\it \
external}, atomic center-of-mass motion with kinetic energy $E_{k}$ have
only recently been explored experimentally \cite
{hood98,ye99a,rempe99,hood00,rempe00a,rempe00b}. In a regime of strong
coupling for the {\it external} degrees of freedom, ${g}_{0}{>E}_{k}/\hbar $%
, a single quantum is sufficient to profoundly alter the atomic
center-of-mass (CM) motion, as an atom moves through a region of spatially
varying coupling coefficient $g(\vec{r})=g_{0}\psi (\vec{r})$ [{\it e.g.},
as arises in the Gaussian mode of a Fabry-Perot cavity, $\psi (\vec{r})$].

Perhaps most strikingly, the spatial variation of the cavity mode can lead
to a confining potential sufficient to trap an atom within the cavity mode
even for a single quantum of excitation of the atom-cavity system, as
first discussed in the work of Refs.\cite{haroche91,walther91}. This is
illustrated in Fig.\ 1, which shows the the possibility for trapping by
excitation to the lower component $|-\rangle $ in
the Jaynes-Cummings
manifold of eigenstates.\ Modifications of the atomic CM dynamics can in
turn significantly alter the cavity field. This situation is very different
from the usual case for trapped atoms or ions in fixed external potentials,
in that here the confining field and the atomic motion can be strongly
interacting, in which case the overall state of the system must be
determined in a self-consistent fashion.

\begin{figure}[h]
\centerline{\psfig{file=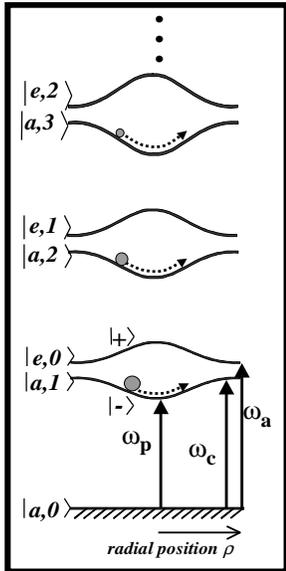,width=1.5in,height=3in}}
\caption{\narrowtext The energy-level diagram for the coupled
  atom-cavity system, as a function of the atom's radial position
  $\rho$. When the atom is near the cavity center, driving at
  frequency $\omega_{p}$ populates the state $|-\rangle$ to trap
  the atom. Here $\omega_{(p,c,a)}=\omega_{(probe,cavity,atom)}$ of
  the text.}
\end{figure}

The experimental requirements to investigate strong coupling for both the
{\it internal} and {\it external} degrees of freedom are stringent [namely, $%
g>({E}_{k}/\hbar ,\gamma ,\kappa )$], and have required the integration of
the techniques of laser cooling and trapping with those of cavity QED, as
was initially achieved in 1996 \cite{mabuchi96} and as illustrated in
Fig.\ 2. Mechanical effects due to strong coupling with single quanta were first
observed in 1998 \cite{hood98}, in an experiment with peak coupling energy $%
\hbar g_{0}\simeq 5$\ mK and with initial atomic kinetic energy $E_{k}\simeq
400\mu $\ K.

Following this theme, two groups recently reported trapping of single
atoms with intracavity fields at the single-photon level, beginning with the
work of Ref.\cite{ye99a} and culminating in that of Refs.\cite
{hood00,rempe00a}. That such trapping might be possible in these experiments
is indicated by the fact that the ratio $R$ of initial atomic kinetic energy
$E_{k}$ to the coherent coupling energy $\hbar g_{0}$, $R\equiv E_{k}/%
\hbar g_{0},$ is less than unity. For the work in Refs.\cite{ye99a,hood00}, $%
R\simeq 0.06$, while for that in Ref.\cite{rempe00a} $R\simeq 0.27$.
Although these ratios are indicative of the possibility of trapping with
single quanta in cavity QED, the actual forces and confining potentials are
somewhat more complex to analyze, as we shall see. Moreover, beyond
providing single-quantum forces sufficient for atomic localization, strong
coupling also means that the presence of one atom can significantly modify
the intracavity field, thereby providing a means to track atomic motion by
way of the light emerging from the cavity.

To understand the basic scheme for trapping of single atoms with single
quanta in cavity QED, consider the energies $\hbar \beta _{\pm }$\ for the
first excited states $|\pm \rangle $ of the atom-cavity system. Along the
radial direction $\rho =\sqrt{y^{2}+z^{2}}$ and for optimal $x$
(standing-wave)\ position, $\beta _{\pm }(\rho )$ has the spatial dependence
indicated in Fig.\ 1, which neglects dissipation. The ground state of the
atom-cavity system is $|a,0\rangle $; the atom is in its ground state $a$,
and there are no photons in the cavity.\ For weak coupling (atom far from
the cavity mode center), the first two excited states are that of one photon
in the cavity and the atom in the ground state, $|a,1\rangle $, and of the
atom in the excited state $e$ with no photons in the cavity, $|e,0\rangle $.
These two states are separated by an energy $\hbar \Delta _{ac}$, where $%
\Delta _{ac}\equiv \omega _{cavity}-\omega _{atom}$ is the detuning between
the ``bare'' (uncoupled) atom and cavity resonances.

\begin{figure}[h]
\caption{\narrowtext Experimental schematic for the case of Hood {\it
    et al}. Atoms are
captured in a magneto-optical trap (MOT), and dropped or launched
through a high-finesse optical 
cavity. A single atom (trace with arrow) transiting the cavity mode alters
the measured transmission of a probe beam through the cavity. In the
experiment of Pinkse {\it et al.}, rubidium atoms are captured in a MOT\
below the cavity and launched upward through it.}
\end{figure}

As an atom enters the cavity along $\rho $ it encounters the spatially
varying mode of the cavity field, and hence a spatially varying interaction
energy $\hbar g(\vec{r})$, given by $g(\vec{r})=g_{0}\cos (k x/
)\exp (-(y^{2}+z^{2})/w_{0}^{2})$ ($k=2 \pi/\lambda$). The bare states
map via this coupling to 
the dressed states $|\pm \rangle $ shown in the figure, with energies 
\begin{equation}
\beta_{\pm }=\frac{\omega _{atom}+\omega _{cavity}}{2}\pm
[g(\vec{r})^{2}+\frac{\Delta _{ac}^{2}}{4}]^{1/2}.
\end{equation}
Our interest is in the state $|-\rangle $; the
spatial dependence of the energy $\hbar \beta _{-}(\vec{r})$ represents a
pseudopotential well that can be selectively populated by our choice of
driving field ${\cal E}_{probe}(t)$ and $\Delta _{probe}$ to trap the atom,
as first suggested by Parkins\cite{scottreference}. The system is
monitored with a weak probe beam as an atom enters the cavity mode;
detection of an atom transit signal triggers an increase in driving strength
to populate the state $|-\rangle $ and trap the atom. Because the
experiments in the optical domain have atomic and cavity decay times ($%
\kappa ^{-1},\gamma ^{-1}$) that are small compared to the time $\tau $ for
motion through the cavity field, the atom-cavity system must be continually
re-excited by way of ${\cal E}_{probe}$, thereby providing an effective
pseudopotential on time scales $\delta t$ such that ($\kappa ^{-1},\gamma
^{-1}$)$\ll \delta t\ll \tau $.

Although a full theory based on the preceding discussion is {\it sufficient}
to provide detailed agreement with the experimental observations of Refs.
\cite{ye99a,hood00,rempe00a} (as we shall show in subsequent sections), it
is reasonable to ask to what extent such a theory based on the interactions
in cavity QED\ is {\it necessary}. In particular, it might well be that the
well-established theory of laser cooling and trapping in free space\cite
{dalibard1985b} could provide an adequate description of the
potentials and heating rates, with the cavity merely providing a
convenient means for
attaining a strong drive field. With respect to the experimental
results of Pinkse 
{\it et al.} (Ref. \cite{rempe00a}), we find that this is in fact largely
the case; there are only small quantitative distinctions between the
free-space theory and the appropriate quantum theory. One interesting
feature to note in this experiment is enhanced cooling of the atomic motion
relative to the parameters of Hood {\it et al.}\cite{hood00}. This effect,
which enables trapping in this parameter regime, arises through 
cavity-mediated cooling \cite{horak1997a,vuletic2000a}. For these
parameters, the average 
localization time from simulations is extended by 75\% relative to the
equivalent free-atom signal; both these times are shorter than the time for
an atom to transit freely through the cavity.

By contrast, in the regime of the experiment of Hood {\it et al.} (Ref.\cite
{hood00}), the cavity QED interactions result in a strong suppression of
dipole heating along the cavity axis relative to the free-space theory,
which has a strong effect on both the duration and character of the observed
atom transits. In the cavity QED setting it becomes possible to create a
potential deep enough to trap an atom without simultaneously introducing
heating rates that cause rapid escape from that potential. For these
parameters, the average experimentally observed localization time is a
factor of 3.5 longer than the equivalent free-atom average. The results of
extensive numerical simulations of trapping times and radial oscillation
frequencies, and their validation by way of comparisons to experimentally
measured distributions, demonstrate the essential role of the single-photon
trapping mechanism in the experiment of Ref.\cite{hood00}. At root is the
distinction between the nonlinear response of an atom in free space and one
strongly coupled to an optical cavity. For these experimental parameters,
the eigenvalue structure of Fig.\ 1 leads to profound differences between
the standard theory of laser cooling and trapping, and the extension of this
theory to the regime of strong coupling in cavity QED.

Note that prior experiments in our group have confirmed that the full
quantum treatment of the one-atom master equation in cavity QED is required
for a description of the dynamics associated with the internal degrees of
freedom for a single atom in an optical cavity in the regime $g>(\gamma
,\kappa )$. These experimental confirmations come by way of measurements of
the nonlinear susceptibility for the coupled system in settings close to
that for the experiment of Ref.\cite{hood00}\cite{hood98,ye99a,mabuchi99}. A
principal goal of this paper is to investigate the extent to which a theory
of atomic motion within the setting of cavity QED is likewise a {\it %
necessary} component in describing the center-of-mass dynamics for the
experiments of Refs.\cite{hood00,rempe00a}.

A second goal is to examine the related question of the extent to which
inferences about atomic motion within the cavity can be drawn from real-time
observations of the cavity field, either via photon counting \cite{rempe00a}
or heterodyne detection \cite{hood00} of the cavity output. The interactions
in cavity QED\ bring an {\it in principle }enhancement in the ability to
sense atomic motion beyond that which is otherwise possible in free space.
Stated more quantitatively, the ability to sense atomic motion within an
optical cavity by way of the transmitted field can be characterized by the
optical information \ $I=\alpha g_{0}^{2}\Delta t / \kappa \equiv
\alpha {\cal R}\Delta t$, which, roughly speaking, is the maximum possible
number of photons that can be collected as signal in time $\Delta t$ with
efficiency $\alpha $\ as an atom transits between a region of optimal
coupling $g_{0}$ and one with $g(\vec{r})\ll g_{0}$. A key enabling aspect
of the experiments in Refs.\cite{hood00,rempe00a} is that ${\cal R=}
g_{0}^{2} / \kappa \gg (\kappa ,\gamma )$, leading to information about
atomic motion at a rate that far exceeds that from either cavity or
spontaneous decay (as in fluorescence imaging). In practice, for detection
strategies employed experimentally, information is extracted at a somewhat
lower rate. For example, in the experiment of Hood et al. \cite{hood00}, the
photon count rate would be $(2.7\times 10^{7}/$s$)$ (including the overall
escape and detection efficiency $\alpha \approx 0.15$), while for the
experiment of Pinkse et al.\cite{rempe00a} it is $(2.2\times 10^{6}/$s$)$
(including an estimated overall escape and detection efficiency $\alpha
\approx 0.11$)\cite{rempe99a}. For time scales $\Delta t\sim 10\mu $\ s as
relevant to the following discussion, atomic motion through the spatially
varying cavity mode leads to variations in the transmitted field that can be
recorded with a high signal-to-noise ratio, namely, a signal of $2.7\times 10^{2}$ photons for
the experiment of Hood et al. and $2.2\times 10^{1}$ for that of Pinkse et
al., where each is calculated for an intracavity field strength of one
photon.

The value of the optical information itself does not tell the complete
story. For cavity QED\ experiments like those considered here, one records
either the sequence of photoelectric counts or the heterodyne current versus
time, from which necessarily only limited inferences about atomic motion can
be drawn. However, if center-of-mass dynamics (i.e., axial and radial
motions) occur on well-separated time scales, then it is reasonable to
suggest that appropriate signal processing techniques could extract
information about these motions from the single time sequence of the
photocurrent $i(t)$. Such processing could presumably occur in real time if $%
\alpha {\cal R}$ is much faster than the rates for radial and axial motion
[e.g., the oscillation frequencies $(f_{r},f_{a})$ in a potential well, with
$f_{r}\ll f_{a}$]. Unfortunately, in neither experiment \cite
{hood00,rempe00a} is $\alpha {\cal R}$ large enough to resolve the axial
dynamics directly, so the task of disentangling the radial and axial motion
signals becomes more difficult, and theoretical simulations of the
experiment become useful in understanding the nature of the observed
transmission signals.

This difficulty arises in the experimental regime of Pinkse et al.\cite
{rempe00a}. For these parameters, axial heating leads to frequent bursts of
large-amplitude motion along the cavity axis, with envelopes extending over
time scales comparable to those for radial motion. Consequently, at
experimental bandwidths (averaging times), both types of motion give rise to
qualitatively similar modulations in the measured transmission signal.
Furthermore, motion in the radial direction has a strong diffusive
component, giving rise to a wide spread of time scales for radial motion. Our
simulations discussed in Sec.\ V suggest that for these parameters,
short-time-scale modulations ($\lesssim 300$\ $\mu$s)\ tend to be
mostly due to 
bandwidth averaging over axial motion, while longer ($\gtrsim 500$\ $\mu$ s)
variations such as presented in Fig.\ 2 of Ref. \cite{rempe00a} typically
reflect radial motion, though these long-time-scale variations are generally
modified in amplitude by the presence of axial motion. Modulations on
intermediate time scales appear ambiguous in their dynamical origin.

By contrast, as shown in Ref.\cite{hood00}, for the parameters of Hood
et al. atoms are well localized along the standing-wave direction throughout
most of the trapping interval, with axial motion giving rise to negligible
signal until finally rapid axial heating leads to atomic escape.
Consequently, observed variations in the photocurrent $i(t)$ are simpler
than those of Ref. \cite{rempe00a}, and directly yield the radial atomic
position. Furthermore, in this experiment the radial oscillation frequency
is large compared to the spontaneous emission heating rate, meaning that the
resulting atomic motion is largely conservative (rather than diffusive)\ in
nature, taking place in a known potential (as demonstrated both
experimentally and by way of numerical simulation). Hence, from $i(t)$ it
becomes possible to make detailed inferences about the radial motion, even
to the point of real-time observations of the anharmonic motion of a single
atom and of the reconstruction of actual atomic trajectories.

The structure of the paper is as follows. Following this introduction,
in Sec.\ II we
present a detailed description of our theoretical model and
its use for the implementation of numerical simulations. Section III\
compares effective potentials and momentum diffusion rates derived for the
two experiments, along with their analogs for the hypothetical case of an
equal-intensity free-space trap. \ These calculations explore the
distinction between quantum and classical, and also give insight into the
nature of atomic motion expected in both experiments. Sample simulated
trajectories are presented for both cases. \ In Sec.\ IV we present
experimental and simulation results for the case of Hood et al., which serve
both to verify the simulations and also to demonstrate important features of
the resulting motion. Sec.\ V\ gives the application of the same tools to
analyze the experiment of Pinkse et al.; we see that standing-wave motion
and diffusive radial motion complicate the correlation between atomic
position and detected field in this case. Finally, axial motion is explored
in more depth, and Fourier analysis of our simulations show that
oscillations of comparable amplitude and frequency should be visible for
both atoms confined (but heated)\ within a well, and atoms skipping along
the standing wave.

\subsection*{Principal findings}

The theoretical treatment and numerical simulation of the motion of a
single atom strongly coupled to an optical 
cavity, as described in Sec.\ II, lead to a surprisingly rich range
of often qualitatively different dynamics. The motion may be
essentially conservative and tightly confined around antinodes of the
standing wave, or essentially dissipative and diffusive and involve
interesting flights between different potential wells of the standing
wave. Indeed we find that the existing experimental results of Hood {\it et
  al.} and Pinkse {\it et al.} exemplify these very different
dynamical regimes. Key features of the atomic motion in both
experimental regimes 
are addressed as follows:

Figures 3--5 and their associated discussion in Sec.\ III
elucidate the nature of the trapping potential and momentum diffusion
in an optical cavity as opposed to a free space standing wave. In
particular we find that, even when the 
atom-cavity system is strongly coupled and driven such that it
has a mean intracavity photon number of roughly $1$, the trapping
potential and momentum diffusion may be only slightly different from
those in a free-space standing wave, and in fact this is the case for
the parameters of Pinkse {\it et al}. On the other hand, for the
parameters of Hood {\it et al.} the usual fluctuations of the dipole
force along the standing wave are suppressed by an order of
magnitude, which to our knowledge represents
qualitatively new physics for optical forces at the single-photon
level within the context of cavity QED.  We show that in the parameter
regime of Pinkse {\it et al.} 
the heating rates are such that the atom could
be expected to gain energy equal to a significant fraction of the
total trapping potential during a single motional oscillation
period for both
axial and radial motion. By this measure the heating rates in the
experiment of Hood 
{\it et al.} are much slower, indicating more nearly conservative
motion, and this could be expected to have a
profound effect on the qualitative nature of the dynamics in the
two experiments.

Figures 6 and 7 and the corresponding text in Sec.\ III\
present simulated transits for both experiments, and discuss the qualitative
features of atomic dynamics in both cases. For the parameter regime of
Hood {\it al.}, conservative radial motion dominates diffusion and
standing-wave motion, with atomic trajectories localized at peaks of a
single standing-wave antinode. Atoms trapped with the mean trapping
time execute several radial orbits. The eventual escape is typically
due to heating 
along the cavity axis. By contrast, for the experiment of Pinkse {\it
  et al.}, a trajectory of typical duration, as in Fig.\ 7(a), does not
experience a complete radial orbit and in fact resembles a scattering
event, with a large contribution from radial diffusion as well. For
these events the observed localization time is comparable to
the time for free flight through the cavity. Axially
the simulations show that  in longer duration transits the atom
frequently skips 
between wells of the standing-wave potential due to repeated heating
and recooling. 

Section IV, with Figs. 8--10, presents a more detailed and
quantitative investigation of trapping and motional dynamics for the
experiment of Hood et al. The ability of our simulations to closely
reproduce the mean trapping times observed in the experiment provides
evidence of their accuracy and utility. As illustrated in Fig.\ 9,
the triggering strategy leads to significant modifications of the
distribution of residence times within the cavity. The essentially conservative
nature of the dynamics and the strong axial confinement make it
possible to confidently ascribe oscillations in the transmitted
intensity to radial motion of the atom. As shown in Fig.\ 10, the
experimentally observed 
oscillations are consistent with the calculated
potential. The conservative nature of the motion is further confirmed
by the separation of orbital periods by angular momentum that is also
apparent in this figure.

Section V, with Figs.\ 11--15, presents a detailed analysis of
trapping and motional dynamics for the experiment of Pinkse et
al. Again, our simulations are sufficient to reproduce the reported
mean localization time. In this case, the triggering strategy leads to
relatively 
minor modifications of the distribution of residence times for an atom
within the cavity. In this case the dissipative nature of the
evolution is significant; essentially no long-term localization is
observed if the 
sign of the friction coefficient is reversed, whereas this has little
effect in the parameter regime of Hood {\it et al.} These largely
dissipative and diffusive motional dynamics are found to have
significant effect on the information about the motion that is
available in the transmitted field. For those events with a long
localization time, the axial motion of the atom
is repeatedly heated and cooled, resulting in slow variations in
envelope of the amplitude of the rapid
oscillations of the transmitted light. The time scale of these
variations is comparable to that for radial motion of the atom. There
are thus no unambiguous signatures for radial motion and only longer
time scale excursions of the atom in the radial potential lead to
variations of the output field that may be confidently ascribed to the
radial motion. Likewise, although information about axial motion is
also available in the output light, we find that it is in
general difficult to distinguish large oscillations in a single well
of the axial potential from free flight over several wells as
attempted in Ref.\ \cite{rempe00a}. 

\section{Theoretical Model and Numerical Simulations}

In this section we outline the derivation from the full quantum-mechanical
master equation of the ``semiclassical model'' for the atomic motion
used in Ref. \cite{hood00}. It turns out that this model is able to
reproduce 
the experimental observations very accurately. Note that here the term
``semiclassical'' refers to approximations with respect to the atomic
center-of-mass motion, and not to the internal degrees of freedom, for
which the full 
quantum character is retained. This situation should not be confused with
the semiclassical theory of cavity QED for which expectation values of field
operators $\hat{O}_{field}$ and atomic operators $\hat{O}_{field}$ are
assumed to factorize, $\langle \hat{O}_{field}\hat{O}_{atom}\rangle =\langle
\hat{O}_{field}\rangle \langle \hat{O}_{atom}\rangle $; no such
approximation is made here. To distinguish these two cases, we introduce the
term {\it quasiclassical} for the case of atomic motion.

The validity of the quasiclassical model depends on a separation of
time scales between the atomic motion and the cavity and internal atomic
dynamics. We adapt the work of Dalibard and Cohen-Tannoudji~\cite
{dalibard1985b} to the situation of a quantized cavity mode. A similar
derivation in the bad-cavity limit appears in~\cite{parkins1996a}. The
details of the derivation are essentially unchanged from free space, since
the terms of the master equation which refer to the dynamics of the cavity
have no explicit dependence on the operators describing the atomic motion.
However, we do find conditions for the validity of the approximation for
this system which depend on the properties of the cavity. Finally, we
describe in more detail the numerical simulations of the resulting model
first presented in Ref.\ \cite{hood00}. These simulations are of the kind
discussed in Refs.\cite{mabuchi98,doherty1997a}.

An analytical calculation of force, momentum diffusion and friction
coefficients for the quasiclassical model of atomic motion in the low
driving limit was derived by Horak and co-workers~\cite
{horak1997a,hechenblaikner1998a}, who found a regime in which the
steady-state temperature scaled as the cavity decay rate. This allows
cooling of the atom below the Doppler limit, so long as the cavity can be
made to have lower loss than the atom. However, the parameters of Refs.\cite
{hood00,rempe00a} are very far from this low driving limit. Hence we employ
numerical techniques based on solving the appropriate master equations by
expansions in terms of Fock states of the cavity field~\cite{tan1999a}. Note
that a very early contribution developed a different theoretical framework
and numerical scheme for calculating the force and friction (but not the
momentum diffusion) of an atom in a cavity (or ``colored vacuum'')~\cite
{mossberg1991a,lewenstein1993a}. Very recently, Vuletic and
Chu~\cite{vuletic2000a} found cavity-mediated cooling in a
slightly different regime to that considered by Horak {\it et al.} 

\subsection{Model of atom-light interaction in a cavity}

The Hamiltonian for a two-level atom interacting with a single mode of the
electromagnetic field in an optical cavity using the electric dipole and
rotating-wave approximations (in the interaction picture with respect to the
laser frequency) is
\begin{eqnarray}
H&=&\frac{\vec{p}^{2}}{2m}+\hbar (\omega _{atom}-\omega _{probe})\sigma
^{\dagger }\sigma +\hbar (\omega _{cavity}-\omega _{probe})a^{\dagger
}a \nonumber \\
&& +\hbar g(\vec{r})(a^{\dagger }\sigma +\sigma ^{\dagger }a)+\hbar \left(
{\cal E}a^{\dagger }+{\cal E}^{*}a\right) .  \label{hamiltonian}
\end{eqnarray}
This is the familiar Jaynes-Cummings Hamiltonian modified to take into
account the external degrees of freedom of the atom and the spatial
variation of the cavity mode. The first term is the kinetic energy of the
atom, and the next two terms are the energy in the internal state of
the atom and 
the cavity excitation. The fourth term describes the position-dependent
interaction of the cavity mode and the atomic dipole. It is important to
note that $\vec{r}$ and $\vec{p}$ are operators. Thus, for example, the
exact strength of the coupling between the atomic internal state and the
cavity field depends on the shape of the atomic wave packet, which is in turn
determined by the mechanical effects of the cavity field. Some implications
of this Hamiltonian are considered in detail by Vernooy and Kimble~\cite
{vernooy1997a}. The Hamiltonian has been written in terms of cavity and
dipole operators that rotate at the frequency of the probe field $\omega
_{probe}$. The real atomic transition (cesium in Ref.\ \cite{hood00}
and rubidium 
in Ref.\ \cite{rempe00a}) in fact involves several degenerate magnetic
sublevels,
but we assume that the cavity is driven by circularly polarized light and
that the atom is optically pumped such that it occupies an effective
two-level system described by the dipole operator $\sigma $ with the
quantization axis along $x.$

Dissipation in the system is due to cavity losses and spontaneous emission.
By treating modes external to the cavity as a heat reservoir at zero
temperature in the Born, Markov, and rotating-wave approximations, it is
possible to derive the standard master equation for the density operator $%
\rho $ of the system \cite{dalibard1985b,carmichael1993a} as
\begin{eqnarray}
\frac{d\rho }{dt} &=&\frac{-i}{\hbar }[H,\rho ]+\kappa (2a\rho a^{\dagger
}-a^{\dagger }a\rho -\rho a^{\dagger }a) \nonumber \\
&&+\frac{3\gamma }{4\pi }\int d^{2}{\bf \hat{k}}S({\bf \hat{k}\cdot \hat{x})}%
\exp (-ik{\bf \hat{k}\cdot r})\sigma \rho \sigma ^{\dagger }\exp (ik{\bf
\hat{k}\cdot r}) \nonumber \\
&& -\gamma (\sigma ^{\dagger }\sigma
\rho +\rho \sigma ^{\dagger }\sigma )
.  \label{mastereqn}
\end{eqnarray}
The third and fourth terms describe the effect of spontaneous emission on
the atomic motion including the momentum kick experienced by the atom as a
result of the spontaneous emission. The unit vector ${\bf \hat{k}}$ is the
direction of an emitted photon. The pattern of dipole radiation is accounted
for by the angular factor $S({\bf \hat{k}\cdot \hat{x})=}\left[ 1+({\bf \hat{%
k}\cdot \hat{x})}^{2}\right] /2$~\cite{javanainen1980a}.

\subsection{Quasiclassical motion of the center of mass}

\label{semiclass}

It is possible to eliminate the internal and cavity dynamics adiabatically
in favor of the slower dynamics of the motional state in parameter regimes
of direct relevance to current experiments. Intuitively, for the
quasiclassical approximation to work, the state of the atom needs to be
sufficiently localized in position and momentum on the scales important to
the problem so that it can be thought of as a classical particle. The
conditions for adiabatically eliminating the internal and cavity dynamics
roughly correspond to this idea. It turns out that it is necessary first
that exchanges of momentum with either the cavity field or by spontaneous
emission into free space should result in momentum kicks that are small
compared with the momentum spread $\Delta p$ of the atomic Wigner function,
thus
\begin{equation}
\varepsilon _{1}\simeq \hbar /\Delta p \ll 1.
\label{crit1}
\end{equation}
For an atom which is in a minimum uncertainty state with respect to the
position-momentum Heisenberg inequality this requires that the state is
localized to better than a wavelength. The atomic motional state will in
general be a mixture allowing the position spread to be broader. However,
this requirement means that the motional state can be thought of as a
probabilistic mixture of pure states localized to within a wavelength, and so
places a limit on the coherence length of the motional state~\cite
{cohentannoudji1992a}. Second it is important that the range of
Doppler shifts of the atom due to its momentum spread is small compared to
the atomic and cavity linewidths, thus
\begin{equation}
\varepsilon _{2}\simeq k\Delta p/m\gamma \simeq k\Delta p/m\kappa
\ll 1.  \label{crit2}
\end{equation}
In this paper it will be assumed that the root-mean-square atomic momentum
obeys this inequality, thus making a low velocity approximation, but the
arguments here can in fact be generalized to arbitrary mean velocities of
the atom~\cite{berg1992a}. The Heisenberg inequality means that this also
requires a minimum position spread of the atom
\begin{equation}
\Delta r\gg \hbar k/m\gamma ,\hbar k/m\kappa .  \label{crit3}
\end{equation}
These criteria are a simple generalization of the situation for laser
cooling in free space which can be imagined as the situation $%
\kappa \rightarrow \infty $. The consistency of these
conditions, which effectively put lower and upper limits on the atomic
momentum spread, requires that
\begin{equation}
\frac{^{\hbar^2 k^{2}/2m}}{\hbar \gamma }\ll 1,\quad \frac{^{\hbar^2
k^{2}/2m}}{\hbar \kappa }\ll 1.  \label{consst}
\end{equation}
The first of these conditions is well known for laser cooling in free
space---the requirement that the recoil energy of the atomic
transition be much 
lower than the Doppler energy, which effectively controls the limiting
temperature of the laser cooling. This condition is well satisfied for heavy
atoms such as cesium and rubidium and the optical transitions employed in
cavity QED experiments considered here. The analogous condition brought
about by the cavity dynamics requires that the recoil energy associated with
exchanging excitation with the cavity field is much smaller than the energy
width of the cavity resonance. Just as the first criterion implies that the
atom still be in resonance with a driving field at its transition frequency
after spontaneously emitting, the second criterion implies that absorbing or
emitting a photon from the cavity will leave the atom near the cavity
resonance. In the experiments of Refs.\ \cite{ye99a,hood00,rempe00a}, $\kappa
\sim \gamma$, so that this second criterion does not place
a stronger restriction on the validity of the approximations than the
free-space limit. However, it is important to note that the design of the
cavity, as well as the atom and transition that are chosen, now has an
effect on the validity of the approximation. It would be possible, for
example, to change the cavity length in such a way that the system moves
from a regime in which the quasiclassical treatment is appropriate into one
in which it is not. In practice for cold atoms cooled to roughly the Doppler
limit ($\Delta p^{2}/2m\sim \hbar \gamma ,\hbar \kappa $) it will be the
case that $\varepsilon _{1}\simeq \varepsilon _{2}\sim \sqrt{\left( \hbar
^{2}k^{2}/2m\right) /\hbar \gamma },\sqrt{\left( \hbar
^{2}k^{2}/2m\right) /\hbar \kappa }$, and so a consistent expansion
should be to equal order in these small parameters.

The derivation of Ref.\ \cite{dalibard1985b} may be applied to our problem, and
proceeds by transforming the master equation~[Eq.\ \ref{mastereqn}] into
an evolution equation for a Wigner operator,
\begin{equation}
W(\vec{r},\vec{p},t)=\frac{1}{h^3}\int d^3
\vec{u} \langle \vec{r}+\frac{1}{2} \vec{u}|\rho
|\vec{r} -\frac{1}{2} \vec{u}\rangle \exp (-i \vec{p}
\cdot \vec{u}/\hbar),
\end{equation}
describing the complete state of
the system. An approximate Fokker-Planck equation for the Wigner function
describing the motional degrees of freedom alone is found by writing this
equation as a Taylor expansion in terms of the small parameters $\varepsilon
_{1}$ and $\varepsilon _{2}$, and truncating that expansion at third order. The
force operator is defined as the gradient of the atom-cavity coupling
\begin{equation}
\vec{F}(\vec{r})=-\hbar g_{0}\vec{\nabla} \psi (\vec{r})(a^{\dagger }\sigma
+\sigma ^{\dagger }a).  \label{fop}
\end{equation}
It is possible to show that the Fokker-Planck equation for the atomic Wigner
function $f$ takes the form
\end{multicols}
\widetext
\begin{equation}
\frac{\partial }{\partial t}f+\frac{\vec{p}}{m}{\bf \cdot }\frac{\partial }{%
\partial \vec{r}}f =-\vec{\phi}(\vec{r}){\bf \cdot }\frac{\partial }{%
\partial \vec{p}}f+\sum_{ij}D_{ij}\frac{\partial ^{2}}{\partial
p_{i}\partial p_{j}}f
+\hbar ^{2}k^{2}\gamma \langle \sigma ^{\dagger }\sigma \rangle _{\rho
_{s}}\sum_{ij}E_{ij}\frac{\partial ^{2}}{\partial p_{i}\partial p_{j}}%
f
 +\sum_{ij}\eta _{ij}\frac{\partial ^{2}}{\partial
p_{i}\partial r_{j}}f +\sum_{ij}\Gamma _{ij}\frac{\partial }{\partial
p_{i}}\left( p_{j}f\right).
\label{eq:fokker}
\end{equation}
The quantities appearing in the Fokker-Planck equation can be calculated
from the master equation for the internal and cavity degrees of freedom
alone that is obtained by setting $\vec{r}$ to some real number value $\vec{r%
}_{0}$, and disregarding the kinetic-energy term. We define $\rho _{s}(\vec{r}%
)$ as the steady state of this master equation, with the steady-state
expectation value of the arbitrary operator $c$ given by $\langle c\rangle
_{\rho _{s}}=$Tr$(c\rho _{s}(\vec{r}))$. The parameters appearing in the
Fokker-Planck equation can then be expressed as follows:
\begin{eqnarray*}
\vec{\phi}(\vec{r}) &=&\text{Tr}\left[ \vec{F}(\vec{r})\rho _{s}(\vec{r}%
)\right],  \\
D_{ij} &=&\int_{0}^{\infty }d\tau \left[ \frac{1}{2}\left\langle F_{i}\left(
\tau \right) F_{j}\left( 0\right) +F_{j}\left( 0\right) F_{j}\left( \tau
\right) \right\rangle _{\rho _{s}}-\phi _{i}\phi _{j}\right] , \\
E_{ij} &=&\frac{3}{8\pi }\int d^{2}\hat{k}S(\hat{k}\cdot \hat{x})\hat{k}_{i}%
\hat{k}_{j}, \\
\eta _{ij} &=&\frac{1}{m}\int_{0}^{\infty }d\tau \tau \left[ \frac{1}{2}%
\left\langle F_{i}\left( \tau \right) F_{j}\left( 0\right) +F_{j}\left(
0\right) F_{j}\left( \tau \right) \right\rangle _{\rho _{s}}-\phi _{i}\phi
_{j}\right] , \\
\Gamma _{ij} &=&\frac{i}{m\hbar }\int_{0}^{\infty }d\tau \tau \left\langle
\left[ F_{i}\left( \tau \right) ,F_{j}\left( 0\right) \right] \right\rangle
_{\rho _{s}}.
\end{eqnarray*}
\begin{multicols}{2}
Simple integrations give $E_{xx}=2/5$ and $E_{yy}=3/10=E_{zz}$ and all other
components of $E$ are zero. Excepting the different definition of the force
operator $\vec{F}$, these are the expressions that can be derived in case of
a free-space light field~\cite{dalibard1985b}. However, it is important to
bear in mind the extra conditions on the validity of the adiabatic
elimination. The master equation~[Eq.\ \ref{mastereqn}] means that the
force expectation values and correlation functions can be very different
from those that are calculated in free space. In practice, the contribution
from the parametric tensor ${\bf \eta }$ is often smaller than that from the
diffusion tensor $D$ by a factor of order $\varepsilon $, and is usually
disregarded in treatments of free-space laser cooling\cite{dalibard1985b}.

Thus, as assumed in earlier work, calculating the quasiclassical motion
of the atom in a cavity field only requires that the force and its
correlation function be evaluated for the full atom-cavity master equation.
Such prior treatments assumed that the atom is motionless; however, they can
be extended to atoms moving at some velocity under the same conditions~\cite
{berg1992a,agarwal1993a}. The diffusion coefficients may be found by first
calculating the correlation functions via the quantum regression theorem and
numerical integration, or directly via matrix-continued fraction techniques~
\cite{berg1992a,agarwal1993a}. A matrix-continued fraction calculation
requires that the field mode be periodic, and as such it only works along the
standing-wave axis of the cavity mode. In directions perpendicular to this,
the calculation of correlations from the master equation is essentially the
only option if the atom is not slowly moving.

\subsection{Stochastic simulations of the quasiclassical model}

It is possible to recast the Fokker-Planck equation of Eq. \ref
{eq:fokker} into a simple set of stochastic equations which describe atomic
trajectories in the cavity field. These equations can be used to gain
intuition about the atomic motion and how it is affected by mechanical
forces. The diffusion and friction tensors can be rewritten using the
definition of the force operator [Eq. (\ref{fop})]
\end{multicols}
\widetext
\begin{mathletters}
\begin{eqnarray}
D &=&\hbar ^{2}g_{0}^{2}\left[ \vec{\nabla} \psi (\vec{r})\right] \left[ \vec{\nabla}
\psi (\vec{r})\right] ^{T}\int_{0}^{\infty }d\tau \left[ \frac{1}{2}%
\left\langle \Phi \left[ \tau \right] \Phi \left[ 0\right] +\Phi \left[
0\right] \Phi \left[ \tau \right] \right\rangle _{\rho _{s}}-\langle \Phi
\rangle _{\rho _{s}}^{2}\right]   \nonumber \\
&=&\hbar ^{2}g_{0}^{2}\xi (\vec{r})\left[ \vec{\nabla} \psi (\vec{r})\right]
\left[ \vec{\nabla} \psi (\vec{r})\right] ^{T}, \\
\Gamma  &=&\frac{i}{m}\hbar g_{0}^{2}\left[ \vec{\nabla} \psi (\vec{r})\right]
\left[ \vec{\nabla} \psi (\vec{r})\right] ^{T}\int_{0}^{\infty }d\tau \tau
\left\langle \left[ \Phi \left[ \tau \right] ,\Phi \left[ 0\right] \right]
\right\rangle _{\rho _{s}}  \nonumber \\
&=&\frac{\hbar g_{0}^{2}}{m}\chi (\vec{r})\left[ \vec{\nabla} \psi (\vec{r}%
)\right] \left[ \vec{\nabla} \psi (\vec{r})\right] ^{T},
\end{eqnarray}
\end{mathletters}
\begin{multicols}{2}
where $\Phi =a^{\dagger }\sigma +\sigma ^{\dagger }a$. Writing the
parameters of the quasiclassical model in this form relies on the
approximation that the atom is slowly moving, namely, that it does not move
a significant fraction of a wavelength during a cavity or atomic lifetime.
Note that the functions $\xi$ and $\chi $ depend on position only through the
coupling $g=g_{0}\psi $. They can be calculated efficiently by finding $%
D_{xx}$ and $\Gamma _{xx}$ using matrix-continued fractions, and then dividing off
the gradient factors. A matrix-continued fraction technique cannot be used
to find the other components of the momentum diffusion or the friction
tensors directly, since the field mode is not periodic across the Gaussian
profile of the mode.

It is now straightforward to convert the Fokker-Planck equation for the
Wigner function into an equivalent set of It\^{o} stochastic differential
equations. The resulting (It\^{o}) equations are \cite{gardiner1985a}
\begin{mathletters}
\begin{eqnarray}
d\vec{x} &=&\frac{1}{m}\vec{p}dt, \\
d\vec{p} &=&-\hbar g_{0}\langle \Phi \rangle \vec{\nabla} \psi dt-\frac{\hbar
g_{0}^{2}}{m}\chi (\vec{r})\left( \vec{p}\cdot \vec{\nabla} \psi \right) \vec{\nabla}
\psi  \nonumber \\
&&+2\hbar g_{0}\sqrt{\xi (\vec{r})}\vec{\nabla} \psi dW_{1}+2\hbar k\gamma
\sqrt{\langle \sigma ^{\dagger }\sigma \rangle }\sqrt{E}d\vec{W.}
\end{eqnarray}
\end{mathletters}
The Wiener increment $dW_{1}$ has the usual properties, in particular $%
dW_{1}^{2}=dt$. The vector $d\vec{W}$ is a vector of three such increments. The
terms in the equation for the momentum are the mean radiative force, its
first-order dependence on momentum, and its fluctuations due to the
atom-cavity system and due to the coupling to free space, respectively. These
equations depend on the quantities $\langle \Phi \rangle ,\chi ,\xi$
and $\langle
\sigma ^{\dagger }\sigma \rangle $, which are functions of position through $%
g$ only. A straightforward simulation of these equations only needs to store
ordered look-up tables of these quantities for given values of $g$, rather
than for all possible values of $\vec{r}$. All of the other quantities that
appear, including $g$, are simple functions of $\vec{r}$ and $\vec{p}$. At each
time step the algorithm searches the look-up table for the current value of $%
g$, starting from the previous value, and reads off the current values of $%
\langle \Phi \rangle ,\chi ,\xi ,\langle \sigma ^{\dagger }\sigma \rangle $.
A linear interpolation for the two closest values of $g$ was used, but more
sophisticated interpolation schemes could be implemented. Since $g$ will not
change by a large amount in any one timestep the search can be very
efficient; a routine from Ref. \cite{press1992a} was used for this. 
In the low-velocity limit of the quasiclassical theory, these stochastic
differential equations describe all the motional dynamics of the atom inside
the cavity. The term proportional to $\eta_{ij}$ leads to correlations
between the atomic position and momentum. The effect of $\eta _{ij}$
is typically small compared to 
friction and diffusion and has been ignored for the moment as is common
practice in free-space standing waves. Terms in the SDE corresponding to the
$\eta $ term in the Fokker-Planck equation could easily be added. This would
mean adding a new noise source which would affect the evolution of the
position as well as the momentum.

\section{Application of the Model to Experimental Regimes}

\subsection{Potentials and heating rates for atomic motion}

The ``quasiclassical'' model discussed in the previous Sec. II can give us a
great deal of information about the nature of the dynamics that may be
expected in the parameter regimes relevant to the experiments of Hood {\em %
et al.} \cite{hood00} and Pinkse {\em et al.} \cite{rempe00a}. In particular
we are interested in whether quantization of the cavity field leads to any
significant change in the dynamics, in the sense of asking whether the
atomic motion is very different in the cavity from what it would be in a
free-space standing wave of the same intensity and geometry as the cavity
mode. Second, we can investigate the nature of the resulting atomic motion
in the cavity field, which can be either predominantly conservative or
significantly diffusive and dissipative, depending on the particular
parameters of interest.

To obtain a feel for the type of atom dynamics expected, effective potentials
and heating rates were calculated for both axial and radial
directions of motion. The effective potential of the atom in the cavity
field may be calculated from the force by
\[
U(\vec{r})=-\int_{0}^{\vec{r}}\vec{F}(\vec{r}^{\prime })\cdot d\vec{r}%
^{\prime }.
\]
The heating rates represent the average increase in the motional energy due
to the momentum diffusion at a given position $\vec{r}$, and may be
calculated from the diffusion tensor according to
\[
\frac{dE}{dt}(\vec{r})=\text{Tr}[D(\vec{r})]/m.
\]
Thus the axial potential at the center of the mode is $U(0,x)=-\int_{0}^{x}%
\vec{F}(0,x^{\prime })dx^{\prime }$ and the associated axial heating rate is
$dE(0,x)/dt=D_{xx}(0,x)/m.$ These quantities along with their radial
equivalents $U(\rho ,0)$ and $dE(\rho ,0)/dt$ are plotted in Fig.\ 3 for
the parameters of Hood et al.\cite{hood00}. The force and momentum diffusion
coefficient for the cavity system were calculated according to the formulas
described above by numerical techniques based on Ref.\
\cite{tan1999a}. The field
state is expanded in terms of number states, and truncated at an appropriate
level and a matrix-continued fraction algorithm is used to calculate
$D$. The axial potentials and heating rates have $\lambda/2=426$\ nm
periodicity inherited from the standing-wave field strength. Observe
that the axial heating rates have minima at both field antinodes and
field nodes. 

\begin{figure}[h]
\centerline{\psfig{file=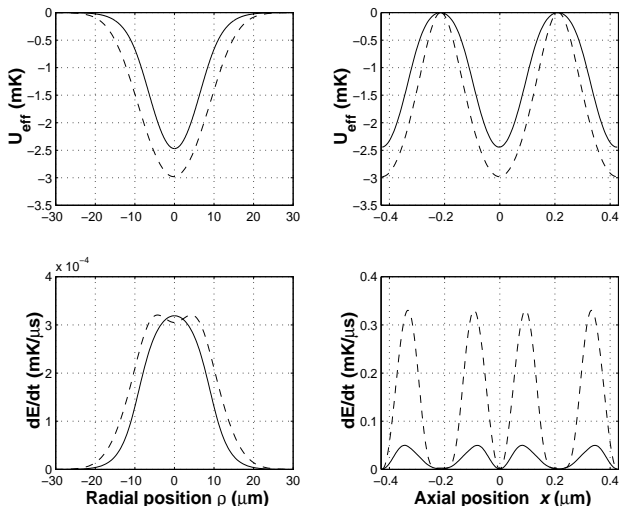,width=3.25in,height=2.67in}}
\caption{\narrowtext Effective potentials $U_{eff}$ and heating rates
  $dE/dt$ in the
radial and axial directions for the experiment of Hood et al.
(solid traces). The cavity field has a Gaussian waist $w_0=14$ $\mu$m in
the radial direction. The axial standing wave has antinodes at
$x=(0,\pm 0.426)\mu m$ and nodes at $x=\pm 0.213\mu m$.  All quantities
are calculated for $\Delta _{probe}/2\pi 
=-125$ MHz and $\Delta _{ac}/2\pi =-47$ MHZ, with an empty cavity
mean-field strength of $\bar{m}=0.3$ photons. For comparison,
corresponding 
quantities for
an equivalent classical free-space trap are shown as dashed
traces. Note that the axial
heating in the cavity trap is tenfold smaller, greatly enhancing the
trap lifetime.}
\end{figure}

The first thing to note is that the axial and radial heating rates are very
different. In the radial direction, heating is dominated by diffusion due to
spontaneous-emission recoils. Axially, however, the reactive or dipole
fluctuation component of the diffusion dominates. This is because the
reactive component is proportional to the gradient of the field squared,
which is much larger for the axial direction where variations are greater
(by a factor of $2\pi w_{0}/\lambda $). This contribution also has the
property that it does not saturate with the atomic response.

It is already clear that it should be possible to trap individual atoms,
since the potential depth of roughly 2.5 mK is greater than the initial
energy of the atoms in the experiment (around 0.46 mK) and the heating rate
in the radial potential is relatively slow. Over 50 $\mu $s (a time scale over
which the atomic motion is strongly affected by the potential) the total
heating will typically still be small compared to the depth of the
potential. However, the importance of the quantum character of the relevant
fields or phenomena is not ensured by the statement that trapping occurs
with a mean field strength of $\bar{m}\sim 1$ photon, since this is
trivially the 
case in an equivalent free-space volume for a field of the same intensity as
that inside the cavity.

\begin{figure}[h]
\centerline{\psfig{file=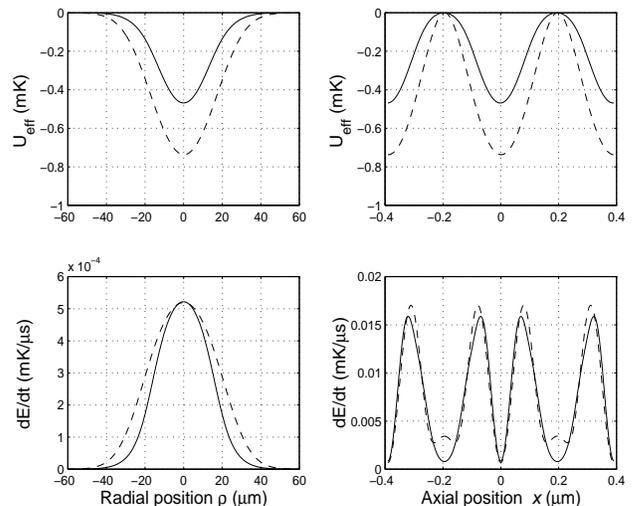,width=3.25in,height=2.67in}}
\caption{\narrowtext Effective potentials $U_{eff}$\ and heating rates
  $dE/dt$\ in 
the radial and axial directions for the experiment of Pinkse et al.
 (solid traces). The cavity field has a Gaussian waist $w_0=29$ $\mu$m in
the radial direction. The axial standing wave has antinodes at
$x=(0,\pm 0.390)$ $\mu$m and nodes at $x=\pm 0.195$ $\mu$m. All quantities
 are calculated for $\Delta 
_{probe}/2\pi =-45$ MHz and $\ \Delta _{ac}/2\pi =-40$ MHz, with an
empty cavity 
photon number $\bar{n}=0.9$. For comparison, corresponding quantities for an
equivalent classical free-space trap are shown as dashed traces. Note
that the
potential depths and heating rates are comparable in the cavity QED\ and
free-space cases.}
\end{figure}

In order to see whether a full quantum description of the atom-cavity is
necessary in order explain observed effects, Fig.\ 3 also shows the values
calculated for an atom in an equivalent free-space standing wave, calculated
by standard techniques\cite{cohentannoudji1992a}. This free-space standing
wave has the same geometry as the cavity mode, and the same peak field
strength $g_{0}|\langle a\rangle |^{2}(0,0).$ The detuning between the
free-space field and the atom is chosen to be $\Delta_{probe}$. Perhaps
surprisingly, the only large difference between the two models is in the
axial heating rate, where a strong suppression of the axial heating is seen
in the quantum calculation. This suppression is an effect of the quantized
nature of the intracavity field. The self-consistent coupling of the cavity
field and atomic position (in a semiclassical sense)\ cannot explain this
suppression; in fact, by itself this coupling would lead to an increase in
diffusion over the free-space case, since the atomic motion within the
cavity induces steeper gradients in the field. The suppression\ of diffusion
is then evidence that it is necessary to use a fully quantum description,
and speak of single photons rather than classical fields for these
experimental parameters. As discussed in Ref.\ \cite{hood00}, this
suppression of 
the axial heating was essential for the trapping of atoms in the cavity.
Thus for these experimental parameters, the eigenvalue structure of Fig.\ 1
leads to profound differences between the standard theory of laser cooling
and trapping and the extension of this theory to the regime of strong
coupling in cavity QED.

\begin{figure}[h]
\centerline{\psfig{file=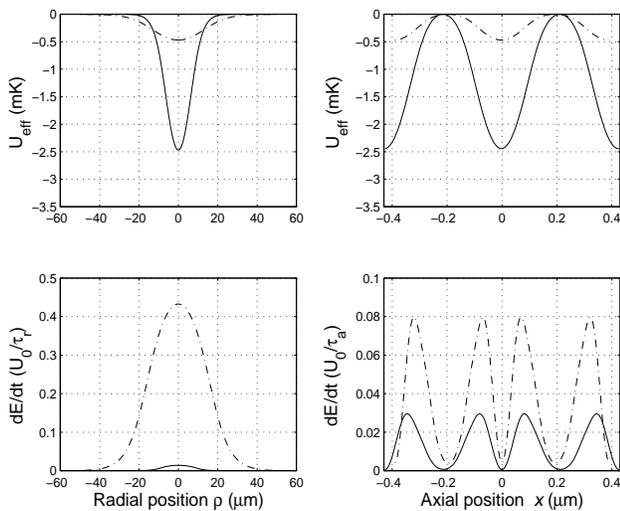,width=3.25in,height=2.67in}}
\caption{\narrowtext Comparison of effective potentials and heating rates in the
  experiments
of Hood et al. (solid line) and Pinkse et al.
(dash-dotted line). Heating rates are shown in units of trap depths
per harmonic 
oscillation period (in the appropriate trap dimension), providing a direct
measure of the degree to which oscillatory motion can be expected to be
conservative in nature. Note that differences in $w_0$ and $\lambda$
between the two experiments lead to quite different radial widths and
slightly different axial periodicities for the quantities plotted. }
\end{figure}

By way of comparison, the same quantities are plotted for the parameters
relevant to Pinkse {\em et al.}\cite{rempe00a} in Fig.\ 4.\cite
{foothetvscount}. The smaller value of $g_{0}$ in this experiment leads to a
smaller effective potential, since the spatial gradients of the dressed
state energy levels (which lead to the potential) are proportional to $g_{0}$%
. More importantly, the diffusion values
calculated from the full quantum model discussed above are now little
different from those of the equivalent free-space standing wave. This lack
of a clear difference in potentials or diffusion indicates that the
quantized nature of the field is not required to explain the radial trapping
observed in Ref.\ \cite{rempe00a}. Note that the resulting axial
heating rates are 
essentially the same as those of Ref.\ \cite{hood00} in absolute magnitude;
however, in Ref. \cite{hood00} the potential was made deeper {\it %
without} the expected corresponding\ increase in diffusion. For the
parameters of Ref.\ \cite{rempe00a} one additional interesting feature
appears---enhanced cooling of the atom motion relative to the
parameters of 
Ref.\ \cite
{hood00}. This arises through cavity-mediated cooling
\cite{horak1997a,vuletic2000a},
and as we shall see, has an important effect on the axial dynamics of atoms
in the experiment of \cite{rempe00a}.

We now wish to use these potentials and heating rates to gain an intuitive
understanding of the character of atomic motion that we would expect to
observe in each case. In particular, we are interested in exploring the
degree to which the atomic motion in the potential can be close to
conservative motion, or likewise the degree to which it could be dominated
by diffusion.

The time scales of relevance to the conservative motion may be characterized
by the period associated with small-amplitude oscillations in the bottom of
the axial ($\tau _{a}=1/f_{a}$) and radial ($\tau _{r}=1/f_{r}$) potential
wells. If the energy changes only by a small fraction (relative to the the
total well depth $U_{0}$) on this time scale, motion will be nearly
conservative. Fig.\ 5 plots the potentials and heating rates for the two
cases in this new set of scaled units; heating rates are expressed as an
energy increase per oscillation period, as a fraction of $U_{0}$ (note as
the atom heats and explores the anharmonicity of the potential, this only
lengthens the period of oscillation). Interestingly, we see a clear
qualitative difference in the nature of the atomic motional dynamics. For
the parameters of Hood et al., in the radial plane spontaneous emission only
gives small perturbations to the energy over the time scale of single orbits,
and motion is nearly conservative. We note that this low level of diffusion
enabled the reconstructions of single-atom trajectories in Ref.\ \cite{hood00},
for which the small changes in angular momentum could be accurately tracked.
A quite different regime is found for the parameters of Pinkse {\em et al., }%
where the radial atomic motion is strongly affected by heating from
spontaneous emission kicks. Here an average atom gains an energy of nearly
half the well depth in what would be a radial orbit time, adding a large
diffusive component to the motion. This same scaling shows that the axial
heating rate is also much more rapid on the scale of the potential in
Ref.\ \cite
{rempe00a}, which suggests that the atom will more quickly escape its
confinement near an antinode and begin to skip along the standing
wave.
The qualitative understanding of the atomic motion gained here is borne out
by the simulations of Refs.\ \cite{hood00} and \cite{rempe00a}, and is
explored in more detail in the simulations to follow.

\subsection{Simulated transits}

Simulations of the kind described in Sec.\ II were performed for the
parameters of the two experiments, and individual instances of these
simulations give insight into the dynamics of the motion---for example, the
relative significance of conservative or dissipative dynamics---and the
correlation between atomic motion and the cavity field state, which is in
turn measured by detection of the output field. Ensembles of these
trajectories provide the statistics of the motion described by the
Fokker-Planck equation [Eq.\ \ref{eq:fokker}] which may then be used to
provide histograms of transit times to compare to the experimental data or
to test reconstruction algorithms for the motion. In order to approximate
the experiment as closely as possible, some effort was made to match the
detailed experimental conditions. The two general considerations were to
reasonably accurately estimate the initial distribution of atomic positions
and momenta for atoms and to consider detection noise and bandwidth when
simulating the feedback switching of the probe laser power.

For each trajectory in the simulations, initial atomic position and momentum
values were drawn from a probability distribution, which was chosen to
correspond to the cloud of atoms following laser cooling and then free fall
\cite{hood00} or launching by an atomic fountain \cite{rempe00a} to the
cavity mirrors. In the simulations, all the atoms started in a horizontal
plane $1$ $\frac{3}{4}$ mode waists above or below the center of the
cavity mode, where mechanical effects on the atom are negligible. Since the
MOT from which the atoms are falling or rising has dimensions much larger
than the cavity mode, the initial position in the axial direction was chosen
from a flat distribution over the cavity mode, and the initial position along
the $y$ axis was also chosen from a flat distribution over 1
$\frac{1}{2}$ mode
waists on either side of the mode center---this distance could be modified
but atoms that are far out in the mode radially do not typically cause large
increases in the cavity transmission, and therefore do not trigger the
feedback. The velocity of the atom along the cavity axis is limited by the
fact that it must not hit one of the mirrors while falling toward the
cavity, and this was also chosen from a flat distribution where the speed was
not more than 0.46 cm/s for the cavity of Hood et al.\cite{hood00}. Although
the two experiments have rather different geometries, we estimate that this
consideration leads to a very similar limiting velocity for motion along the
axis. In the experiment of Pinkse et al.\cite{rempe00a}, we used 0.4
cm/s.
The velocity along the $z$ axis was chosen from a Gaussian distribution
appropriate to the temperature of the MOT ($\sim 20$ $\mu K$) after
polarization gradient cooling. For \cite{hood00} the velocities in the
vertical direction were chosen by calculating as appropriate for an atom
falling freely from the MOT (the MOT\ is situated 3.2mm above the mode with
a spatial extent of standard deviation 0.6mm). Thus atoms arriving at the
cavity axis have a mean vertical velocity $\bar{v}=25$cm$/$s. Some of these
parameters such as the height, size and temperature of the initial MOT\ are
not precisely known for the experiment, so that some consideration of the
variation of the histograms and other features of the resulting simulations
has been made although no systematic optimization in order to obtain the best
agreement has been undertaken. In Ref.\ \cite{rempe00a} the mean
initial vertical
velocity of atoms entering the cavity is $20$ cm/s. This speed is very much
less than the mean velocity imparted to the atoms by the pushing beam which
launches them from the MOT $25$ cm below, and as a result the atoms are all
near the top of their trajectories. Simple kinematical calculations show that
the resulting distribution of velocities should be rather broad compared to
the mean. In the absence of more detailed information about the MOT\
temperature and spatial size and the strength of the pushing beam we choose
the initial vertical velocity distribution to be a Gaussian of mean 20
cm/s
and standard deviation 10 cm/s---this leads to a distribution of trapping
times with a mean that matches the mean reported in Ref.\ \cite{rempe00a}. Each
trajectory proceeds until the atom is either a greater radial distance from
the center of the mode than it started from, or it has moved
sufficiently far in 
the axial direction that it would hit one of the cavity mirrors.

The detection and triggering are modeled as follows. In the parameter range
in which the ``quasiclassical'' model is valid, the cavity field comes to
equilibrium with the atomic position on a time scale much faster than the
atomic motion itself, and thus the light transmitted through the cavity (over
bandwidths of the order of tens to hundreds of kilohertz) is associated with
the atomic motion. At each point in the simulation the intracavity field and
intensity expectation values are stored in order to record for each
trajectory a noiseless and infinite-bandwidth trace. In practice,
experimental traces will look like filtered and noisy versions of these
traces. As an atom enters the cavity mode, a weak driving field is present
for probing. \ In order to model the triggering step, the field intensity $%
\langle a^{\dagger }a\rangle $ or field amplitude modulus squared $|\langle
a\rangle |^{2}$ is averaged over a time equal to the bandwidth of the detection in
the case of heterodyne detection as in Ref.\ \cite{hood00}, or over the time
windows in which photocounts are binned in the case of direct photodetection
as in Ref.\ \cite{rempe00a}. A random number with the appropriate variance to
represent the shot noise is added and the total is compared with some
predecided level---if the transmission exceeds this level the probe laser
beam is increased in strength in order to attempt to trap the atom. In the
case of Ref.\ \cite{hood00} the trigger level is $|\langle a\rangle|
^{2}=$ 0.32, 
the averaging time is 9 $\mu $s, and there is a 2-$\mu $s delay
between triggering and changing the driving laser power. For the
experimental bandwidth of 100 kHz, the appropriate noise has standard
deviation 0.05 at a transmitted signal of 0.32. These parameters are chosen
so as to match as closely as possible the conditions of the experiment. The
same procedure is followed for simulations of the parameters \cite{rempe00a}%
. Although the exact triggering protocol is not described there, we assumed
that counts over a period of $10$ $\mu $s were used to decide whether or
not to trigger and the noise was chosen to be consistent with the reported
photon count rate of $2\times 10^{6}$s$^{-1}$ \cite{rempe99a}.

Examples of such trajectories are plotted for the parameters of Ref.\
\cite{hood00} 
in Fig.\ 6 and for those of Ref.\ \cite{rempe00a} in Fig.\ 7. The
chosen trajectories 
range in length from the experimentally reported mean transit time upward,
and are chosen because they show typical features of the dynamics in each
case. It is clear that the two experiments are in quite different parameter
regimes, as already indicated by the relative sizes of the potentials
and heating rates.

For the parameters of Ref.\ \cite{hood00}, the atoms orbit in a radial
plane; some 
have nearly circular and some very eccentric orbits. The motion along the
axial direction is usually well localized near an antinode of the
standing wave, where the axial heating rate is small. This localization
occurs because atoms are channeled into the antinodes by the weak potential
associated with the initial probing field, which slowly begins to affect an
atom as it falls across the mode waist during the detection stage of the
experiment. However the strong axial heating that is present away from the
antinodes means that once an atom begins to heat axially, it suffers a burst
of heating (over several hundred microseconds), which leads to its loss from
the potential well associated with a single antinode of the field.
Frequently the atom leaves an axial potential well when it is radially far
from the center of the cavity mode, since in this case the axial potential
becomes weaker. Note that the mean transit time in Ref.\ \cite{hood00}
corresponds 
to $\sim3.5$ radial orbits around the center of the cavity mode, so
transits with multiple oscillations are frequently observed.
In Ref.\ \cite
{rempe00a} the radial oscillation frequency is slower, so an atom of mean
transit time does not in fact make a complete rotation about the mode
center. The radial motion in this case is also visibly more stochastic in
nature, as a result of the relatively faster spontaneous emission momentum
diffusion discussed above.

Another interesting difference between the two parameter regimes is, as
suggested in Ref.\ \cite{rempe00a}, the relative importance of atomic motion along
the standing wave as opposed to oscillations around a single antinode. In
the case of Ref.\ \cite{rempe00a}, long, strongly trapped transits
almost always 
involve intervals when an atom is skipping along the standing wave, as well
as intervals when it is oscillating in an individual well. By contrast, for
the parameters of Ref.\ \cite{hood00}, only a few percent of
trajectories involve 
skipping during times in which the atom is trapped, and this is usually
associated with movement over one or two wells with the atom falling back
into the adjacent or a nearby well. This happens so quickly that it does not
affect the radial motion in practice, or lead to a detectable signal in the
output light, so that these rare events of skipping do not affect the
reconstructions of Ref.\ \cite{hood00}.

\begin{figure}[p]
\centerline{\psfig{file=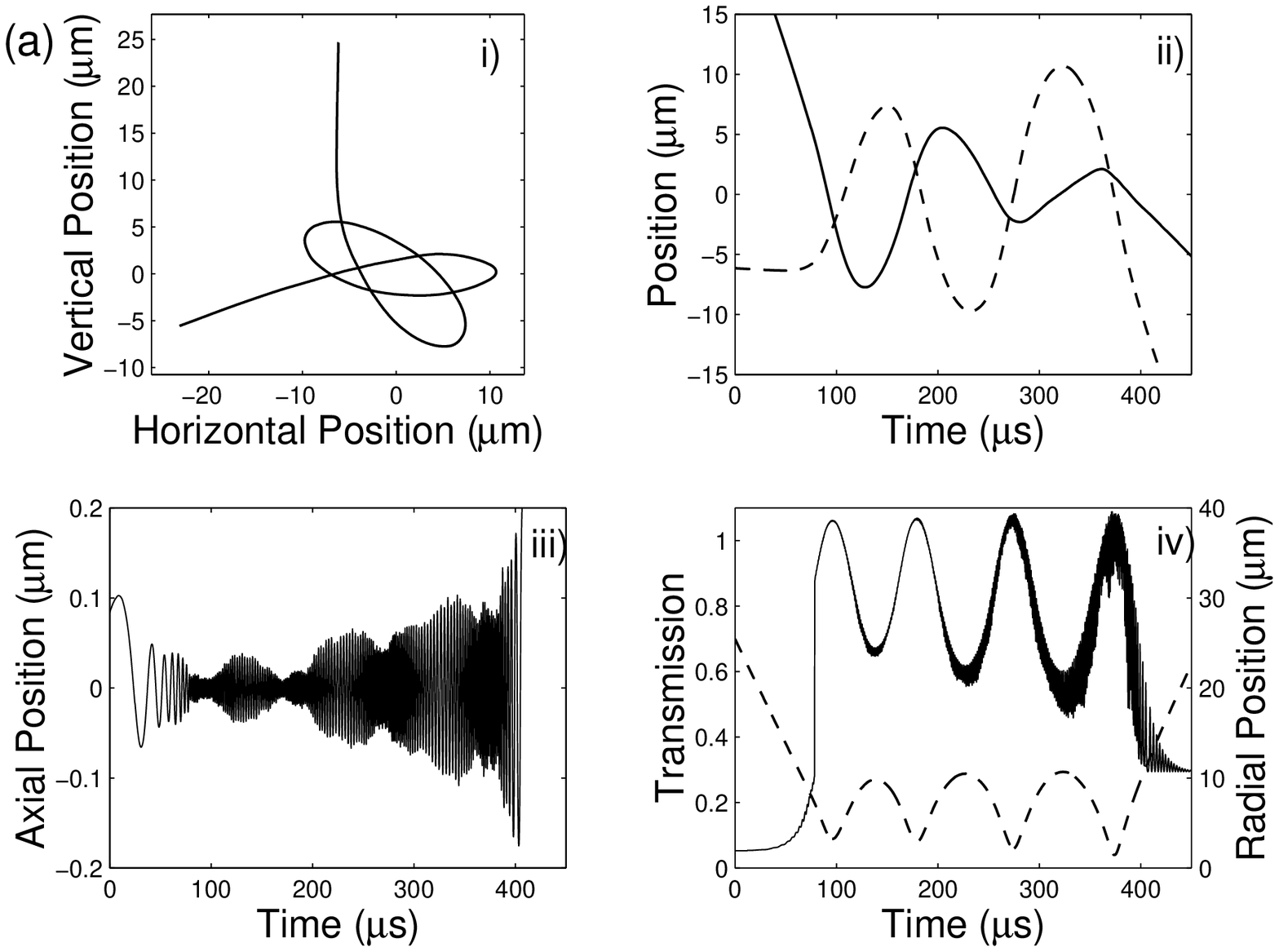,width=2.925in,height=2.34in}}
\centerline{\psfig{file=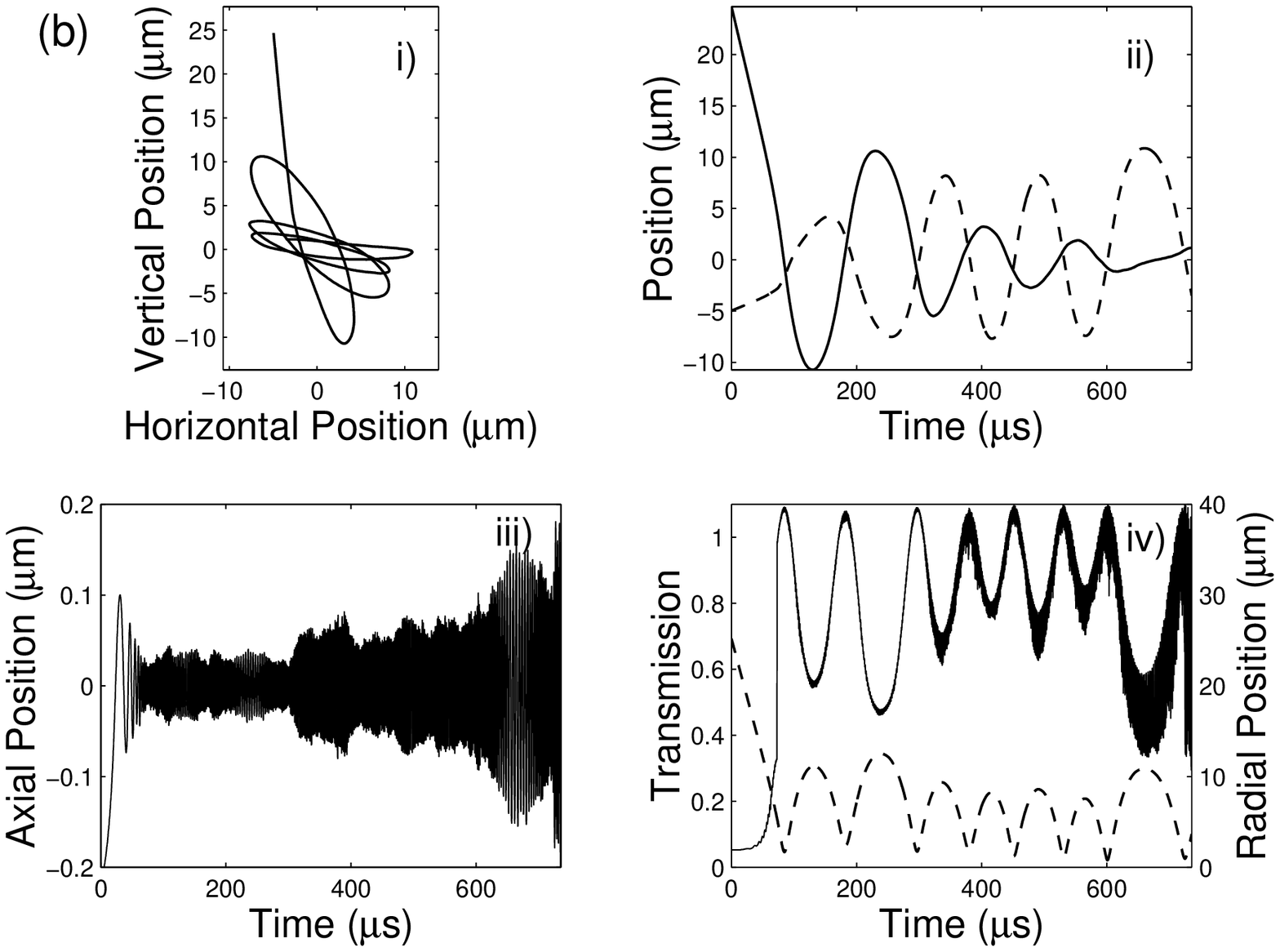,width=2.925in,height=2.34in}}
\centerline{\psfig{file=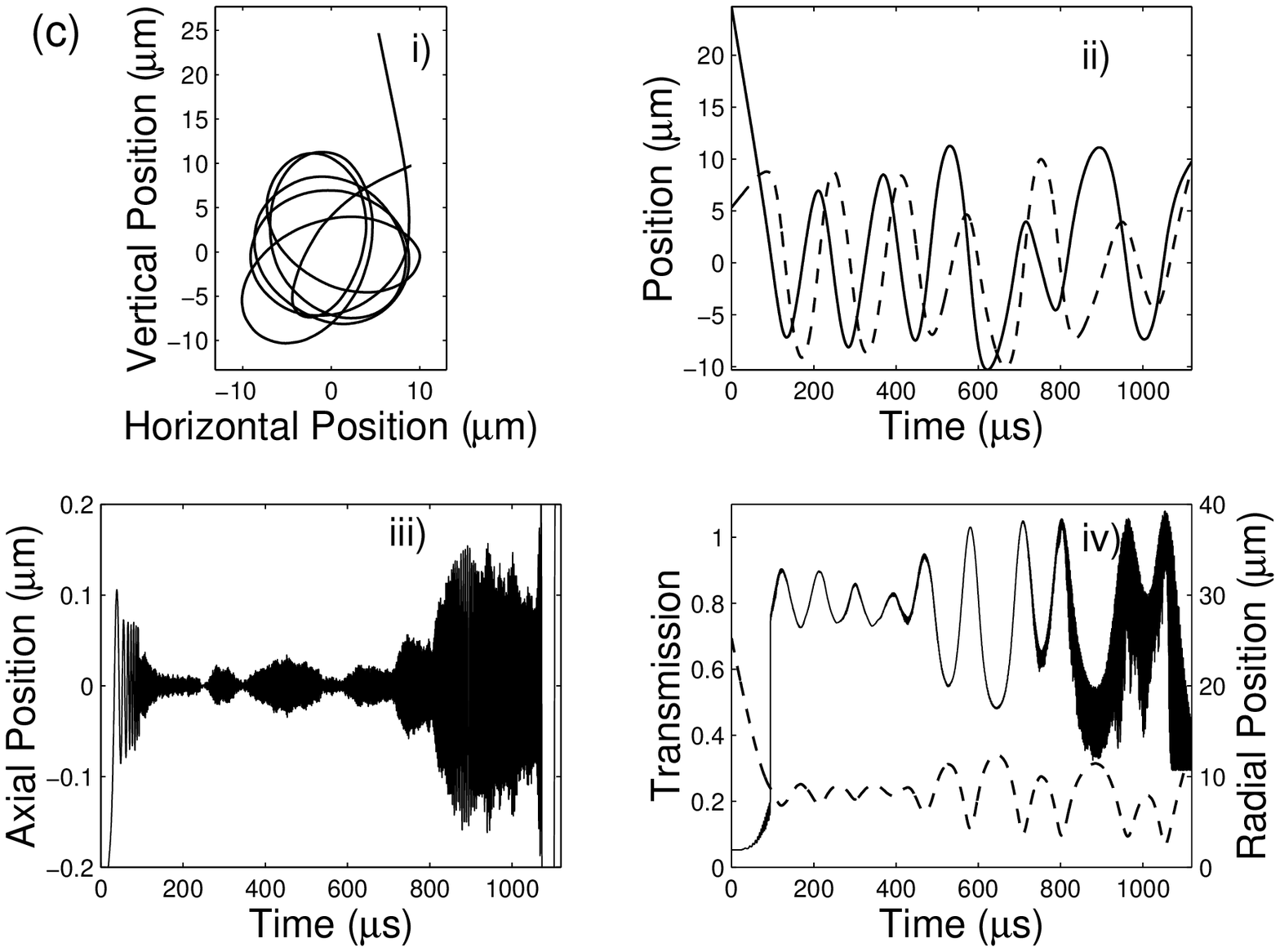,width=2.925in,height=2.34in}}
\caption{\narrowtext Typical trajectories from simulations of the experiment of
  Hood {\em et
al.}\ as described in the text. The driving parameters are $\Delta
_{probe}/2\pi  
=-125$ MHz and $\Delta _{ac}/2\pi =-47$ MHz, with an empty cavity
mean-field strength of $\bar{m}=0.3$ photons. The trajectories have
transit
durations of (a) $345$ $\mu$s, (b) $680$ $\mu$s, and (c) $1032$
$\mu$s. This is one,
two, and three mean transit times respectively. (i) The radial trajectory of
the atom; the $z$ position is plotted against the $y$ position. (ii)
The $y$ position
(dashed line ) and $z$ position (solid line) are plotted as a function
of time. (iii) The
axial position, where zero is an antinode of the cavity field. (iv) The
noiseless infinite-bandwidth transmission $\bar{m}$ (solid line) and the radial
distance from the center of the mode (dashed line).}
\end{figure}
\begin{figure}[p]
\centerline{\psfig{file=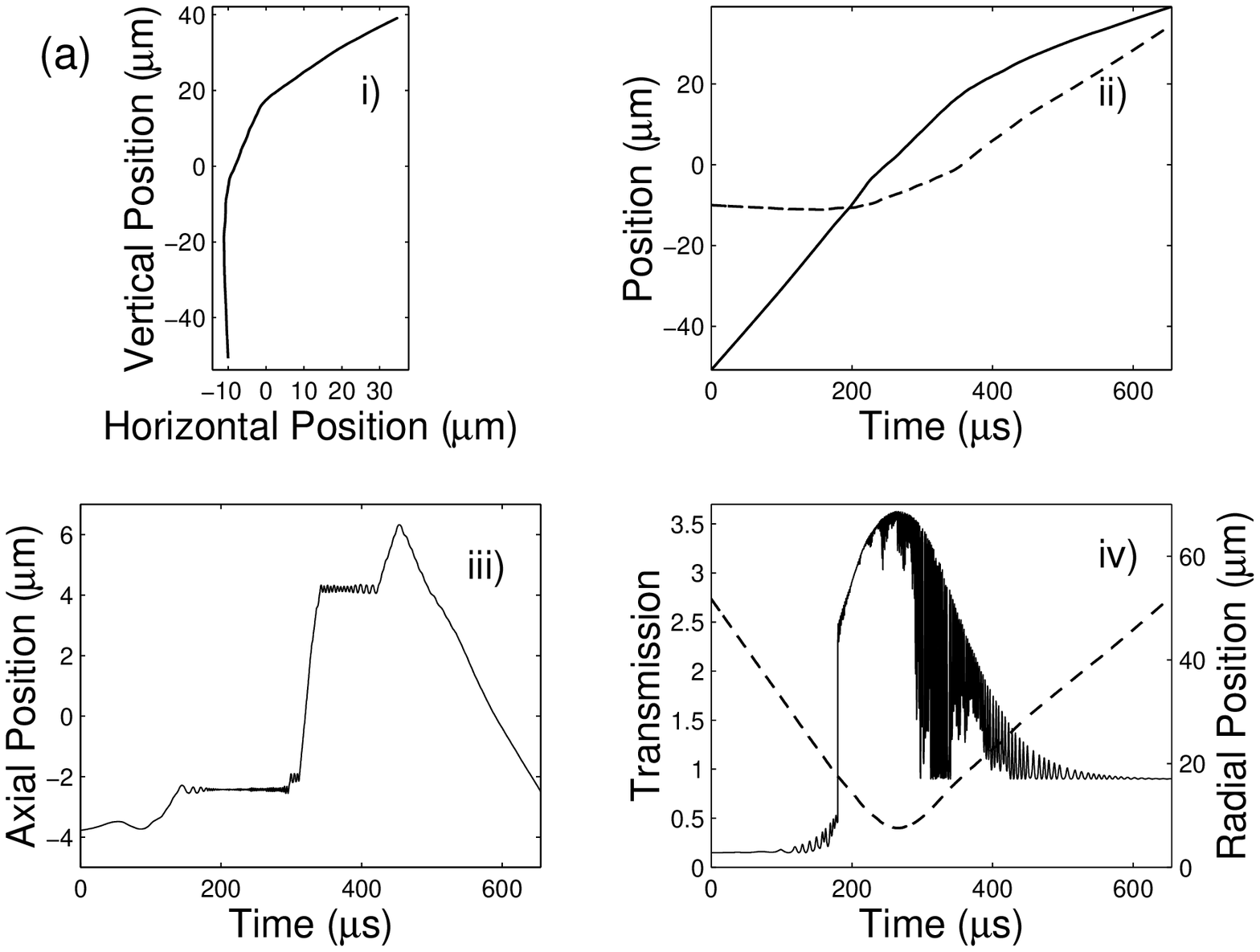,width=2.925in,height=2.34in}}
\centerline{\psfig{file=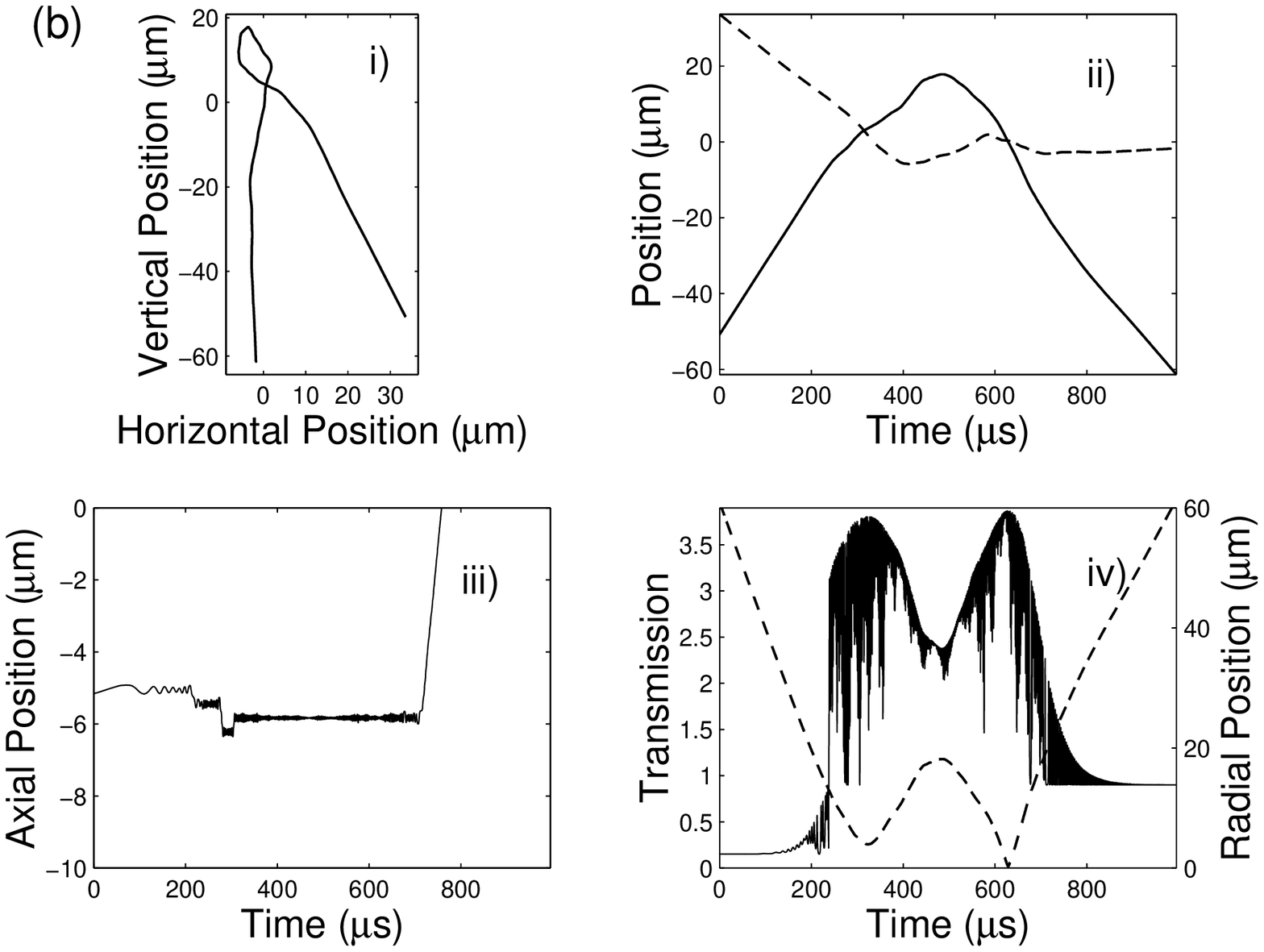,width=2.925in,height=2.34in}}
\centerline{\psfig{file=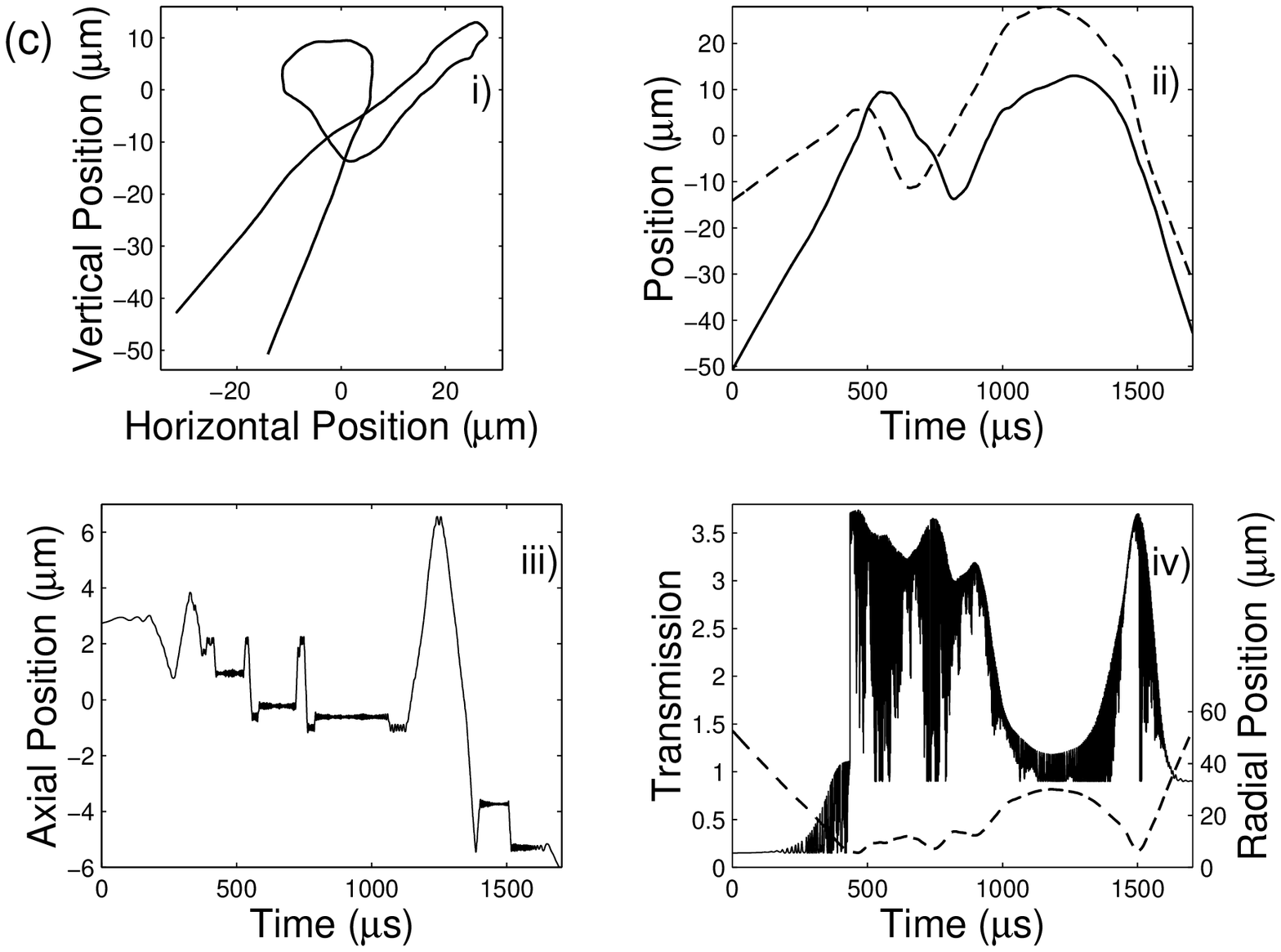,width=2.925in,height=2.34in}}
\caption{\narrowtext Typical trajectories from simulations of the experiment of
  Pinkse {\em
et al.}\ as described in the text. The driving parameters are $\Delta 
_{probe}/2\pi =-40$ MHz and $\ \Delta _{ac}/2\pi =-35$ MHz, with empty cavity
photon number $\bar{n}=0.9$. The trajectories have
transit durations of (a) $247$ $\mu$s, (b) $514$ $\mu$s, and c) $1358$
$\mu$s. The
experimentally reported mean transit time is $250$ $\mu$s. (i) The radial
trajectory of the atom; the $z$ position is plotted against the $y$
position. (ii) The
$y$ position (dashed line) and $z$ position (solid line) are plotted
as a function of 
time. (iii) The axial position, where zero is the mean axial position over the
transit. (iv) The noiseless infinite-bandwidth transmission $\bar{n}$
(solid line)
and the radial distance from the center of the mode (dashed line).}
\end{figure}

As noted in Ref.\ \cite{hood00}, the axial
motion often becomes more significant at the end of a transit and as the
atom is leaving the mode, which leads to atoms skipping a well in perhaps as
many as one in five cases at the end of the transit. We find from the
simulations that in Ref.\ \cite{rempe00a}, the first escape time from an axial
potential well for an atom initially localized near an antinode is
sufficiently short compared to the mean trapping time that skipping along
the wells almost always takes place. On the other hand, the first escape
time is of the order of several times the mean trapping time for the
parameters of Ref.\ \cite{hood00}, so skipping between standing wells is
correspondingly rare.

It is interesting to note that the friction coefficient for the parameters of
Pinkse {\em et al.} is much more significant than for the experiment of
Hood {\em et al.}, and plays an important role in the axial motion of the
atom. As in the trajectories shown here it is a feature of essentially every
trajectory for the parameters of Ref.\ \cite{rempe00a} that the atom
spends time
in potential wells associated with several different antinodes of the field.
However, we performed simulations with the sign of the friction coefficient
reversed, and found that no more than a few percent of trajectories were
recaptured in a second well after having begun to skip along the standing
wave. Clearly the dissipative nature of the motion is an integral feature of
the dynamics in this regime, and in particular it enables the atoms to fall
back into axial potential wells after escape due to the rapid heating in
that dimension.

\section{Simulation Results for the Experiment of Hood et al.}

Having presented the theoretical basis underlying the simulated atom
trajectories, in this section we present results of these simulations and
their comparison with experimental results as reported in Ref.\cite{hood00}.
We generate a set of simulated trajectories for the parameters $%
(g_{0},\gamma ,\kappa )=2\pi (110,2.6,14.2)$ MHz with detuning parameters $%
\Delta _{ac}=\omega _{cavity}-\omega _{atom}=-2\pi \times 47$ MHz and $\Delta
_{probe}=\omega _{probe}-\omega _{atom}=-2\pi \times 125$ MHz. In
correspondence with the experimental protocol, the initial pretriggering
level of the driving laser gives a 0.05-photon mean-field strength in the
empty cavity; when this level rises to 0.32 photons indicating the presence
of an atom, we trigger a sixfold increase in the driving strength to a
trapping level of a 0.3-photon empty-cavity mean field strength. A close
correspondence between theory and experiment is obtained for these results,
demonstrating the relevance of this theoretical model to the physics of the
actual experiment. In addition, both theoretical and experimental results
exhibit features which are relevant to building up a picture of the nature
of the single-atom, single-photon trapping and atomic dynamics, both
qualitatively and quantitatively.

We begin by presenting the qualitative similarity of experimental and
simulated atom transit signals, as observed via detection of cavity
transmission as a function of time. Fig.\ 8 shows two sample experimental
transits [(a) and (b)] and two sample simulated transits [(c) and (d)]. For
the simulated 
transits, traces of the corresponding radial and axial motion are also
shown. Transmission is shown here as $\bar{m}=|\langle a\rangle |^{2}$, as
is appropriate for the balanced heterodyne detection of Ref.\cite{hood00}.
In the case of the simulated results, the simulated transmission signal has
been filtered down to the experimental detection bandwidth of 100 kHz, and
both technical noise and shot noise have been added\cite{footnoise}. The
transmission signal thus processed can be seen to lose some of the clarity
with which it reflects the full atomic dynamics, in comparison to the
transmission traces of Fig.\ 6. In particular, the experimental detection
bandwidth is much slower than the time scale for axial oscillation in the
confining potential, so that observed transmission signals are averaged over
the fast variation in $g$ caused by these axial oscillations.\
The observed
maximum transmission should therefore be lowered relative to theoretical
predictions, by an amount dependent on the amplitude of typical axial
motion. Thus this finite-bandwidth effect allows for an experimental
estimation of the axial confinement of a typical transit. Such a procedure
gives an estimate of confinement within $\sim 70$ $nm$ of an antinode, in good
agreement with simulation results which suggest typical confinement within $%
\sim 50$ $nm$. It is important to note that while such tight confinement
appears typical over the duration of a trajectory, atoms commonly undergo
rapid diffusive heating near the end of their confinement lifetime, which
leads to their escape in a majority of cases.

%%\vspace{4.33in}
\begin{figure}[tbh]
\caption{\narrowtext (a) and (b) Examples of atom transits, i.e.,
  cavity transmission as  a function
of time as an atom passes through the cavity field for the experiment of
Hood et al. Solid traces show atoms trapped using the
triggering method described, with an $\bar{m}\simeq 1$ photon peak field
strength. For comparison, an untriggered (untrapped) atom transit is shown
in the dashed trace. For these traces, the parameters are those of
Fig.\ 3. The empty-cavity 0.3-photon mean-field strength is indicated by the
horizontal dashed line. (c) and (d)Theoretical simulation of atom transits for
the same $\Delta _{probe}$ and $\Delta _{ac}$. Shot noise and technical
noise have been added to the transmission signals, which have also been
filtered to experimental bandwidth. Other traces show the radial
(dashed line)
and axial (solid line) motion of the atom. Motion along $x$, the standing-wave
direction, has been multiplied by 10 to be visible on the plot. Note
that the
atom is very tightly confined in $x$ until rapid heating in this direction
causes the atom to escape.}
\end{figure}

\subsection{Trapping lifetimes}

From the entire set of experimental and simulated trajectories like those of
Fig.\ 8, it is possible to investigate some quantitative aspects of the
trapping dynamics. First we focus on the trap lifetimes produced by the
triggered-trapping scheme. Figures 9(a) and 9(b) show histograms of
experimental
transit times for untrapped atoms and for atoms trapped by means of the
triggered-trapping strategy. Transit durations are determined from the
experimental data by recording the time interval during which the
transmission signal is clearly distinguishable from the empty-cavity
transmission level, in the presence of experimental noise. Since the
signal-to-noise ratio for observing transits depends on the specific probe
parameters, one must be careful to compare untriggered and triggered
transits observed with the same detunings and intracavity field strengths.
The sole difference must be that in the untriggered case, the empty cavity
field is set at a constant strength so that the atom falls through the
effective potential, whereas in the triggered case the field begins at a
lower level and is only turned up once the atom enters the cavity, thus
confining the atom.\ For example, Fig.\ 8(a) shows sample untriggered
(dashed) and triggered (solid) transit signals which correspond to one
another in this way.

In Fig.\ 9 the difference in transit lifetimes between triggered (b) and
untriggered (a) cases is immediately striking. For their initial fall
velocity of $\bar{v}=25$ cm/s, atoms have a free-fall
time of $\sim110$ $\mu$s across the cavity waist $2w_{0}=2(14.06$ $\mu$m). As
discussed above, the duration of observed transits is limited by the
signal-to-noise ratio, which provides a slightly more restrictive cut on transit
durations, so the untriggered data set shows a mean duration of $92$ $\mu$s.
In contrast, when the triggered-trapping strategy is employed, the mean
trapping lifetime is $340$ $\mu$s. The dispersion about the mean
likewise changes drastically from $75$ $\mu$s in the untriggered case
to $240$ $\mu$s in the triggered case. These results represent a clear
signature of the 
trapping of single atoms with single photons via this method. In this
setting, atoms have been observed to remain trapped in the cavity field for
as long as $1.9$ ms.

The corresponding theoretical histograms are shown in Figs.\ 9(c) and
9(d) for the 
untriggered and triggered cases. The start of the transit is taken to be the
time at which an atom could be distinguished in the cavity given the signal
to noise, and the final time is taken to be the last point at which the
transmission dropped to within the noise of the transmission with no atom.
This definition accounts for the fact that as atoms move out in the radial
direction the transmission often drops to around the free space value, but
returns again to some large value over the time scales of the atomic motion.
These levels were chosen to duplicate as closely as possible the protocol
for deciding transit times for the experimental data.

The simulated transit set shows a mean trapping time of $96$ $\mu$s in the
untriggered case and $383$ $\mu$s in the triggered case and dispersions
of $84$ and $240$ $\mu$s, respectively. This result is in
good agreement with the experimental results when statistical errors and
uncertainties in the initial MOT parameters are taken into account. The
agreement between experimental and simulated trap lifetimes, in both mean
and distribution, gives an indication of the validity of the theoretically
calculated trapping potential and diffusive forces on the atom. The 3.5-fold
increase in observed lifetimes due to trapping is made possible by the
cavity QED\ interaction, which allows creation of a deep trapping potential
without correspondingly large diffusion as in the free-space case.

\begin{figure}[tbh]
\centerline{\psfig{file=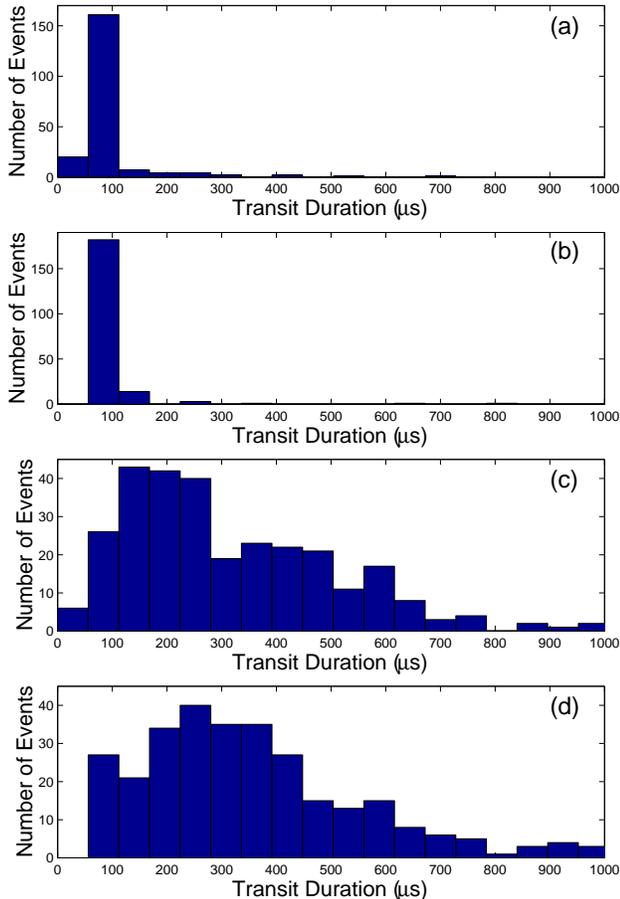,width=3.25in,height=4.77in}}
\caption{\narrowtext Observed atomic transit durations for untriggered and triggered cases,
with the parameters of Fig.\ 3. (a) and (c) Experimental
data show a mean observation time of 92 $\mu$s in the untriggered case (a)
and 340 $\mu$s in the triggered case (c), indicating the significant
trapping effect. For comparison, the free flight time across the
cavity waist is
110 $\mu$s. (b) and (d) The simulated transit set shows a mean of 96
$\mu s$ for 
the untriggered case (b) and 383 $\mu s$ for the triggered case (d), in good
agreement with experiment.}
\end{figure}

\subsection{Oscillations and radial motion}

We now turn to a more detailed investigation of the dynamics of motion
experienced by a trapped atom. As we have seen, the transmission signal for
a single trapped atom exhibits large variations over time which may be
tentatively identified with atomic motion in the radial (Gaussian)
dimensions of the cavity field. Thus, for example, the highest transmission
occurs when the atom passes closest to the cavity axis, $\rho =0$. To
determine the validity of such an identification, we examine the periods of
observed oscillation in the transmission signal. The calculated effective
potential is approximately Gaussian in the radial dimension, so a
one-dimensional conservative-motion model predicts periods as a function of
oscillation amplitude in this anharmonic effective potential well. Referring
to the sample transits of Fig.\ 8, one does indeed note a trend toward
large modulations with long periods and smaller modulations with shorter
periods. To quantify this observation, we plot period $P$ versus the
amplitude $A$ for individual oscillations, where $A\equiv
2[(H_{1}+H_{2})/2-H_{c}]/(H_{1}+H_{2})$, with $\{H_{1},H_{2},H_{c}\}$ as
indicated in Fig.\ 8. 

\begin{figure}[t]
\centerline{\psfig{file=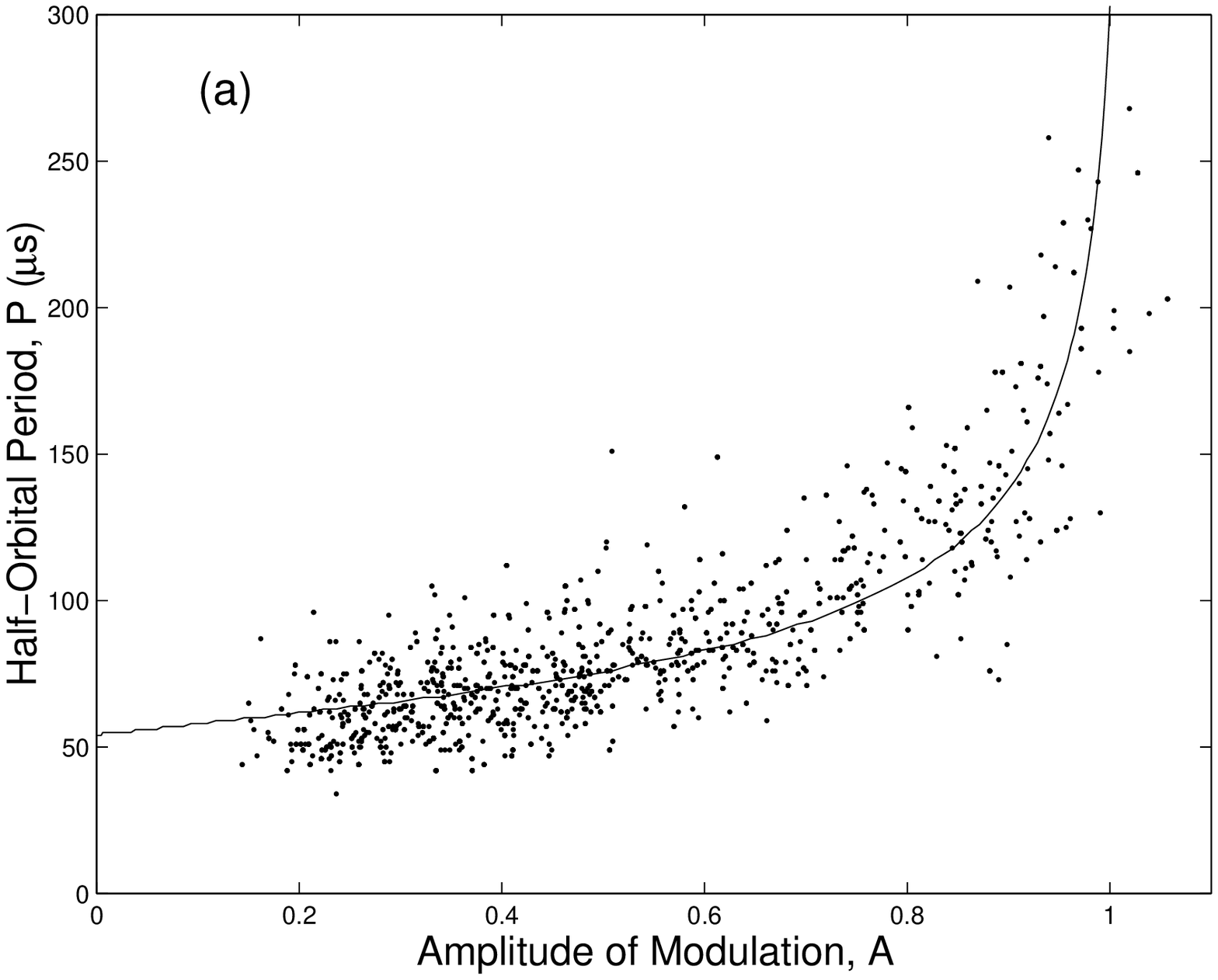,width=3.25in,height=2.67in}}
\centerline{\psfig{file=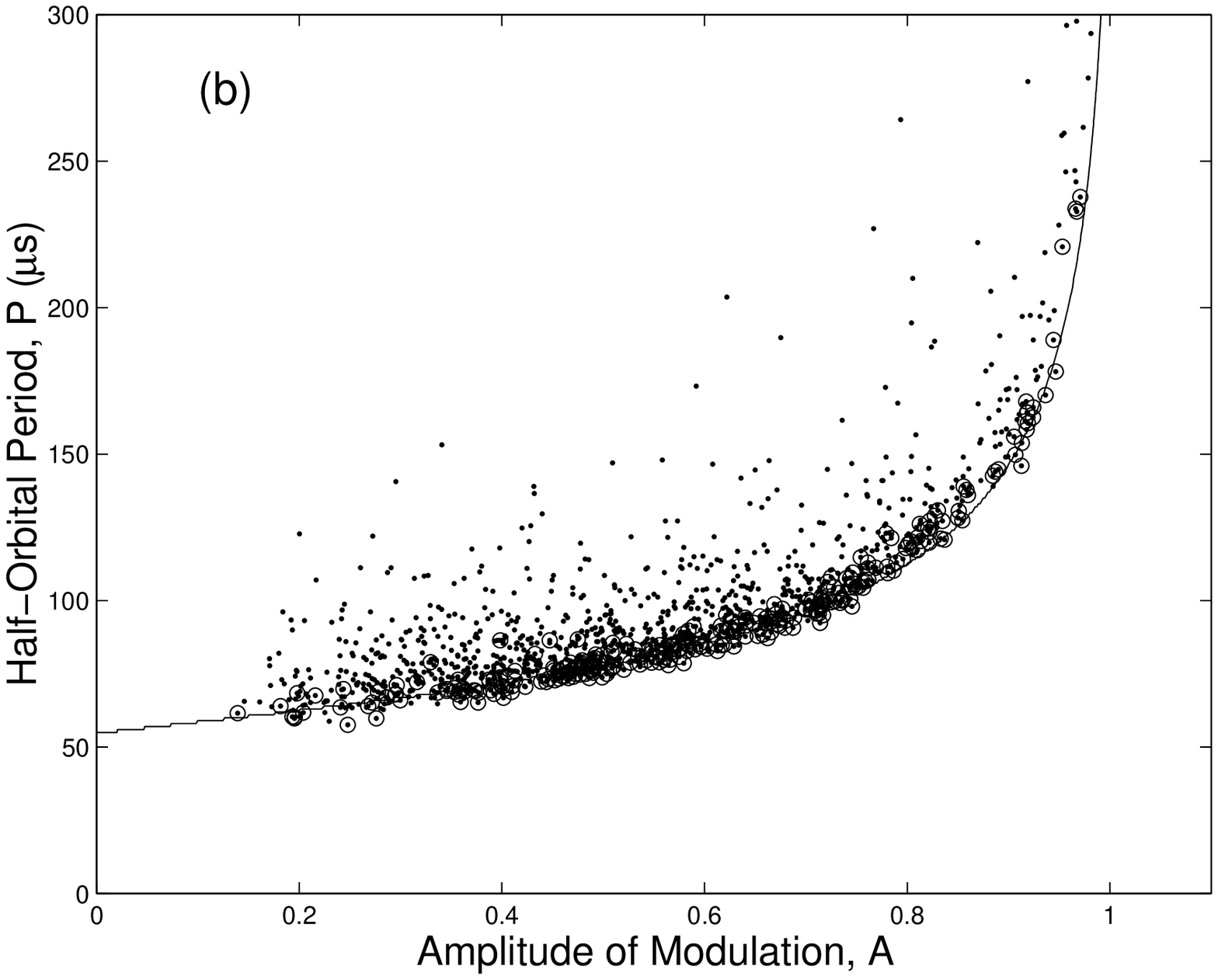,width=3.25in,height=2.67in}}
\caption{\narrowtext (a) For experimental transmission data of Hood et
  al., the modulation
period is shown as a function of amplitude. If modulations in
transmission are 
tentatively identified with radial atomic motion, their expected period is
half that of the radial motion. \ The solid curve gives calculated period
vs amplitude based on this assumption and on one-dimensional motion in the
effective potential $U(\rho ,0)$ of Fig.\ 3. (b) Corresponding plot for
simulated transmission data. \ Points with lowest underlying atomic angular
momentum are plotted with circles; separation by angular momentum reflects
the conservative nature of atomic dynamics on time scales comparable to a
radial period.}
\end{figure}

Figure 10(a) shows the experimental data plotted along with the calculated
curve for one-dimensional motion in the effective potential $U(\rho ,0)$\
(see Fig.\ 3), for the same parameters as Fig.\ 8. (This is a different
data set from that presented in Fig.\ 4. of Ref \cite{hood00}.)\ Note that
since an atom approaches the cavity axis $\rho =0$ twice over the course of
one orbital period, the predicted period for oscillations in the
transmission signal is half the period of the underlying atomic motion.
Experimental data clearly map out this calculated curve for radial atomic
motion, demonstrating that oscillations in the observed cavity transmission
do indeed reflect radial position of an atom as it varies over time within
the trap. The agreement also indicates the quantitative correctness of the
theoretical model for the radial potential depth and spatial profile. Note
that the comparison is absolute with no adjustable parameters.

The same analysis may be performed for transmission oscillations in the set
of simulated transits, yielding the plot of Fig.\ 10(b). This plot again
shows agreement with the calculated curve, with some spread away from the
line.\ For simulated transits, it is possible to turn to the underlying
atomic position record to determine an angular momentum for the atom during
a given oscillation. Thus the oscillation data of Fig.\ 10(b) are plotted by
atomic angular momentum, where lower angular momentum data points are shown
with circles.{\it \ }A separation by angular momentum is clearly evident,
with lower angular momentum points most closely following the calculated
one-dimensional (and thus zero angular momentum) curve. This separation,
while it may seem expected, is in fact a non-trivial indication that angular
momentum is a valid quantity for the atomic motion over the course of an
oscillation period. Since the atomic motion is not in fact conservative, but
is also influenced by random (diffusive) forces, a separation by angular
momentum can only be expected to occur if the effect of diffusive forces is
sufficiently small over the time scale of an orbit in the conservative
potential. The plots of Fig.\ 3 provide an initial indication that this is
indeed the case for these parameters, and this idea is borne out by the
current investigation. Confidence in the relatively small effect of
diffusion over a single orbital period is crucial in the reconstruction of
two-dimensional atomic trajectories as in Ref.\cite{hood00}.

\section{Simulation Results for the Experiment of Pinkse et al.}

Having provided a validation of our capabilities for numerical simulation by
way of the results of Sec. IV, we next apply this formalism to
the experiment reported in Ref.\ \cite{rempe00a}. At the outset, we note that
the various approximations discussed in Sec.\ II related to the derivation
of this quasiclassical model are satisfied to a better degree for this
experiment than for the experiment of Ref.\ \cite{hood00}. Hence we expect
that the correspondence between the simulations and experiment should be at
least of the quality as in the preceding section.

Our starting point is the generation of a large set of simulated
trajectories for the parameters reported in Ref.\ \cite{rempe00a}, namely, $%
(g_{0},\gamma ,\kappa )=2\pi (16,3,1.4)$ MHz with detuning parameters $\Delta
_{ac}=\omega _{cavity}-\omega _{atom}=-2\pi \times 35$ MHz and $\Delta
_{probe}=\omega _{probe}-\omega _{atom}=-2\pi \times 40$ MHz. The initial
pretriggering level of the driving laser gives a 0.15-photon mean intensity
in the empty cavity; when this level rises to 0.85 photons, indicating the
presence of an atom, we trigger an increase in the driving strength to a
trapping level of 0.9-photon empty-cavity intensity. \ These criteria are
intended to follow the parameters indicated in Figs.\ 2 and 3 of Ref.\cite
{rempe00a}. Note that for the cavity geometry of this experiment, the time
for an atom to transit freely through the cavity mode in the absence of any
light forces is $\tau _{0}=2w_{0}/\bar{v}=290$ $\mu $s, where as before we take
twice the cavity waist $w_{0}$ as a measure of the transverse dimension of
the cavity.

\subsection{Histograms of transit durations}

From the set of such simulated trajectories ($\sim400$ in this particular
case), we can construct histograms for the number of events as a function of
total transit signal duration. Following the experimental protocol of Ref.%
\cite{rempe00a}, which employed photon counting, we base this analysis upon
the intracavity photon number $\bar{n}=\langle a^{\dagger }a\rangle $ rather
than$|\langle a\rangle |^{2}$ as in Ref.\cite{hood00}, although this
distinction is not critical to any of the following considerations.

\begin{figure}[tbp]
\centerline{\psfig{file=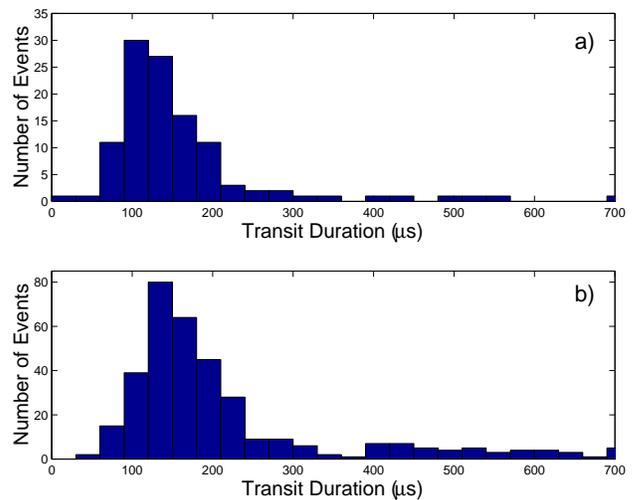,width=3.25in,height=2.62in}}
\caption{\narrowtext Simulated atomic transit durations for untriggered and
  triggered cases,
with the parameters of Pinkse et al., as in Fig.\ 7. (a) The
untriggered transit set shows a mean observation time of 160 $\mu$s. (b)
The triggered transit set shows mean duration 280 $\mu$s, in good agreement
with the experimentally quoted mean of 250$\pm $50$\mu$s. \ For comparison,
free-fall time across the cavity waist is 290 $\mu$s.}
\end{figure}
The resulting histograms for the experiment of Ref.\cite{rempe00a} are
displayed in Fig.\ 11 for the cases of untriggered and triggered
trajectories. As in the discussion of Fig.\ 9, the external drive strengths
are set to be equal for this comparison to provide equal detectability for
an atom passing through the cavity mode. Detection with lower external drive
strength gives a lower signal-to-noise ratio for atom detection, which
results in 
detected transit durations much shorter than the actual passage time through
the cavity (which is of order $\tau _{0}=2w_{0}/\bar{v}$), as for example in
Fig.\ 2(a) of Ref.\cite{rempe00a}.

In support of the validity of our simulations for the experiment of Pinkse
et al. (including the initial atomic velocity and position distribution and
the triggering conditions), note that the mean of $280$ $\mu $s for the
histogram in the triggered case of Fig.\ 11(b) corresponds quite well with
that quoted in Ref.\cite{rempe00a}, namely, $\bar{\tau}_{\exp }=250\pm 
50$ $\mu $s. Further, the histograms in Fig.\ 11 exhibit an extension of the
mean transit duration from $160$ $\mu $s for the case of no triggering in (a)
to $280$ $\mu $s with triggering in (b), in support of the claim of trapping in
Ref.\cite{rempe00a}. The dispersion of
events around the mean is quite large in both cases, $161$ $\mu$ s in
the untriggered set and $282$  
$\mu$s in the triggered set. The increase in the mean is largely
associated with an increase in 
the number of events in the range 200-300 $\mu$s, as well as in the number of
rare events much longer than the mean duration. Once again we note that the
dissipative nature of the dynamics plays a crucial role in the observed
motion for the experiment of Pinkse {\em et al}. A histogram of transit
durations calculated with the sign of the friction coefficient reversed has
a lower mean than that of transits with no triggering.

However, it is certainly worth noting that the observed ``average trapping
time'' $\bar{\tau}_{\exp }=250\pm 50$ $\mu $s quoted in Ref. \cite
{rempe00a}, as well as the corresponding mean time from our simulations, are
smaller than the time $\tau _{0}=290$ $\mu $s for an atom to transit freely
through the cavity mode. Additionally, even in the case of no triggering,
there is already a significant number of events with similar long duration
to those in (b) with triggering. Such events arise from the relatively large
contribution of diffusion-driven fluctuations whereby an atom randomly loses
a large fraction of its initial kinetic energy as it enters the cavity. That
such fluctuations play a critical role should already be clear from the
plots of the confining potentials and diffusion coefficients in Fig.\ 4.

\subsection{Radial motion}

Trapping dynamics can also be explored if atomic oscillation in the trapping
potential can be directly observed. Certainly the observations presented in
Fig.\ 10 make this case for the experiment of Ref.\cite{hood00}, with the
observed oscillation frequencies found to be in good quantitative agreement
with those computed directly from the anharmonic potential without
adjustable parameters and with the results of the numerical simulations.

Towards the goal of constructing a similar plot for the parameters of Ref.
\cite{rempe00a}, consider a long-duration transit event such as that in
Fig.\ 7(c). Recall that the output flux from the cavity is given by the
cavity decay rate $2\kappa _{d}$ into the relevant detection channel times
the intracavity photon number, or $I=2\kappa _{d}\bar{n}=2\kappa _{d}\langle
a^{\dagger }a\rangle $, with then the detected count rate found from the
overall propagation and detection efficiency as $R=\xi I$. Of course, in any
actual experiment the full information displayed for the intracavity photon
number $\bar{n}$ is not available because of finite detection efficiencies $%
(\xi <1)$ and the requirement to average over many cavity lifetimes in order
to achieve an acceptable signal-to-noise ratio (roughly for a time such that
$\sqrt{R\delta t}>\gg 1)$.

\begin{figure}[h]
\centerline{\psfig{file=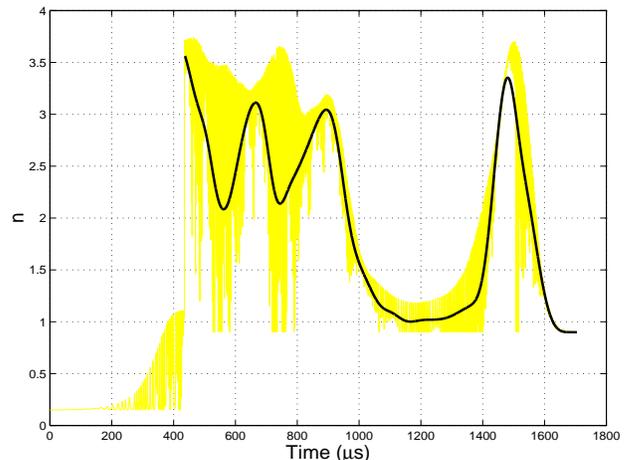,width=3.25in,height=2.44in}}
\caption{\narrowtext Transmission data for the simulated transit of
  Fig.\ 7(c). The full
ideal signal $\bar{n}(t),$ with infinite bandwidth and no
degradation due to cavity escape efficiency or subsequent system losses, is
shown in gray. Slow variations are caused by radial motion while fast
variations reflect axial motion. The black trace results from applying to
this ideal signal a low-pass filter with cutoff $f_{c}=10$ kHz intended to
optimize the visibility of any radial oscillations for frequencies $%
f\lesssim 5$ kHz, where $f_{0}^{(r)}=2.6$ kHz is the orbital frequency for
small-amplitude oscillation near the bottom of the radial potential.
The resulting filtered transmission signal shows 
variations due to both radial motion and axial heating.}
\end{figure}

Rather than attempt a detailed analysis of such effects for the experiment
of Ref.\cite{rempe00a}, here we wish to illustrate several generic effects
that hinder definitive observation of radial oscillations in this regime. We
therefore take {\it the full ideal signal }$\bar{n}(t)${\it \ with no
degradation due to cavity escape efficiency or subsequent system losses}
(which we estimate to be $\kappa _{d} / \kappa \sim 0.17$ and $\xi \sim 0.6$%
 for an overall efficiency of 0.11). As shown in Fig.\ 12, to this ideal
signal we apply a low-pass filter with cutoff $f_{c}=10$ kHz intended to
optimize the visibility of any radial oscillations for frequencies $%
f\lesssim 5$ kHz, where $f_{0}^{(r)}=2.6$ kHz is the orbital frequency for
small-amplitude oscillation near the bottom of the radial
potential. As before, recall that a periodic variation in the
radial coordinate at frequency $f$ results in a variation in $\bar{n}$ at $%
2f $. Precisely such a filtering protocol was implemented for the analysis
in Fig.\ 10, there with $f_{c}=25$ kHz in correspondence to the larger
radial oscillation frequencies ($f_{0}^{(r)}=9.4$ kHz for
Ref.\cite{hood00}) \cite{footfilters}.

Not surprisingly, the frequent and large bursts of axial heating evident for
the simulated trajectories of Fig.\ 7 result in large variations in the
intracavity photon number on time scales set by twice the axial oscillation
frequency $f_{0}^{(a)}\approx 430$ kHz. While these axial oscillations cannot
be directly resolved in the detected counting signal $R(t)$, their envelope
nonetheless leads to variations in $\bar{n}(t)$ and hence $R(t)$ on time
scales comparable to that associated with radial motion (i.e., $1 /
2f_{0}^{(r)}$), as is apparent in Fig.\ 12. Consequently, the low-pass
filtering [or, equivalently, the time averaging over segments in $R(t)$] that
is required experimentally to obtain an acceptable signal-to-noise ratio
gives rise to observed variations in $\bar{n}(t)$ that can arise from either
axial or radial atomic motion. In the particular transit shown in Fig.\ 12,
two apparent variations on time scales $\simeq200$ $\mu$s are introduced by a
filtering of the axial motion, whereas the longer modulation ($\simeq
600-\mu$s duration) does reflect the radial position of the atom. This is
something of a generic feature of the several hundred simulated transits
examined; shorter-time-scale modulations ($\lesssim 300$ $\mu$s) can reflect
either a genuine radial excursion or a filtering of axial motion, whereas
very long period variations (500--600 $\mu$s) are indicative of radial atomic
motions. This simply reflects the fact that the bursts of axial motion tend
to have time scales limited to a few $100$ $\mu$s.

To illustrate these points further, we have constructed a plot of period
versus normalized amplitude of transmission oscillations from our
simulations of the experiment of Pinkse et al.\cite{rempe00a}, with the
result given in Fig.\ 13. We emphasize that the protocol followed is
precisely as for the analysis that led to Fig.\ 10(b) for the experiment of
Hood et al.\cite{hood00} (see also Fig.\ 4 of Ref.\cite{hood00}), with the
exception of the aforementioned reduction in the low-pass cutoff frequency.
In marked contrast to that case, here there is a poor correspondence between
the distribution of orbital periods from the ensemble of simulated
trajectories and the prediction from the potential obtained from Eq.\
(\ref{fop})% 
. Referring to the discussion of Fig.\ 12 above, we note that about 2/3\ of
the points in the 100--300 $\mu$s range result from averaging over axial
motion, whereas for longer-period (P>300 $\mu$s) modulations, 80\%\ of\
the observed points reflect changes in
the radial motion, but with associated transmission amplitude typically
modified by the presence of axial motion. The results of Fig.\ 13 [which
are for the ideal case of $\bar{n}(t)$ without signal degradation due to
finite escape and detection efficiency] suggest that only in restricted
cases can temporal variations in $R(t)$ be attributed to radial motion, and
not instead of (or in addition to) the envelopes of axial heating processes.
Indeed, such effects are well known in the literature, having been
previously discussed for the case of individual atoms falling through the
cavity mode (albeit without triggering or trapping) \cite
{doherty1997a,mabuchi99}. A similar conclusion was reached, namely, that
axial heating processes contaminate the frequency band associated with
radial motion, thereby precluding inferences about radial motion. For the
data presented by Pinkse et al. \cite{rempe00a}, the long ($\simeq
500$ $\mu$s%
)\ time scale of the modulations suggests an assignment of these signals to
radial motion; however, a more detailed characterization of the atom dynamics
over a larger ensemble of transits should yield this more
definitively.

It is also worth noting that the quoted average trapping time{\it \ }$\bar{%
\tau}_{\exp }=250\pm 50$ $\mu $s in Ref.\cite{rempe00a} is itself less
than $1/f_{0}^{(r)}=390$ $\mu $s, which is {\it shortest time} for a
full radial orbit. Hence any conclusion about motion in the radial plane
must necessarily be based upon rare events in the tail of the histograms of
Fig.\ 11. The rare occurrence of these long events is reflected in the
small number of data points in Fig.\ 13, which was constructed from the
same number of simulated transits as Fig.\ 10(b).

\begin{figure}[h]
\centerline{\psfig{file=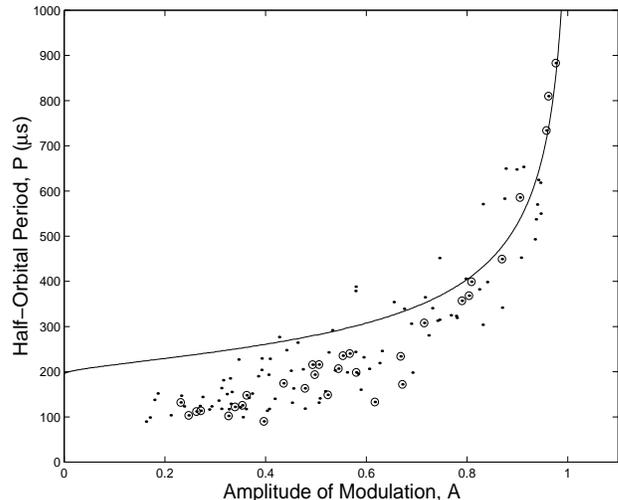,width=3.25in,height=2.67in}}
\caption{\narrowtext For simulated transmission data corresponding to
  the parameters of Pinkse et al.,
the modulation period is shown as a function of amplitude. If modulations in
transmission are tentatively identified with radial atomic motion, their
expected period is half that of the radial motion. The solid curve gives
calculated period vs amplitude based on this assumption and on
one-dimensional motion in the effective potential $U(\rho ,0)$ as in Fig.\ 4.
Points with lowest underlying atomic angular momentum are plotted with
circles. Lack of separation by angular momentum reflects the diffusive
nature of atomic dynamics on time scales comparable to or shorter than one
radial period.}
\end{figure}

\subsection{Axial Motion}

We next turn to analyze motion along the axial direction, and to the
statement of Pinkse et al.\cite{rempe00a} that Fig.\ 4 of Ref. \cite
{rempe00a} ``is direct evidence for the atom moving along the cavity axis,''
as opposed to instances of localization around an antinode for which
``hardly any periodic structure is visible.'' In their analysis, Pinkse et
al. employed a function $g^{(4)}(\epsilon ,\tau ,\epsilon )$, whose intention
is to pick out two-time correlations in intensity, with an enhanced
signal-to-noise ratio of intensity fluctuations by measuring coincidences of
photon pairs. Here we attempt to investigate manifestations of the axial
motion independent of the details of any specific such function by analyzing
$\bar{n}(t)$ directly by way of a windowed fast-Fourier transform (FFT).
More specifically, for each trajectory from a large ensemble from our
simulations, we apply a FFT to the record $\bar{n}(t)$ with a Hanning
window centered at time $t_{i}$ and of total width $25$ $\mu $s, with the
window then offset sequentially to $t_{i+1}=t_{i}+5$ $\mu $s to cover the whole
range of a given atomic trajectory. The window width $25$ $\mu $s is chosen to
be in close correspondence to the record length of $20$ $\mu $s employed by
Pinkse et al. Longer window widths do not qualitatively change the results
of our analysis, while a substantially shorter-duration window leads to a
loss of requisite frequency resolution.

Two examples from an extended set of such transforms are given in Figs.\ 14
and 15. Parts (a) of each of these figures show the mean intracavity photon
number $\bar{n}(t)$, the axial coordinate $x(t)$, and a contour plot of the
windowed FFT ${\cal N}_{t_{i}}(\Omega )$ for a single atomic trajectory for
the parameters of Ref.\cite{rempe00a}. Here ${\cal N}_{t_{i}}(\Omega )$ is
the windowed FFT of $\bar{n}(t)$ over the entire duration of the trajectory,
with $t_{i}=t_{0}+i \times 5$ $\mu $s. Parts (b) of Figs.\ 14 and 15 compare
${\cal N}% 
_{t_{i}}(\Omega )$ for two particular values of $t_{i}$, namely, at a time $%
t_{flight}$ corresponding to the midst of a flight of the atom over several
antinodes of the intracavity standing wave (i.e., variations in axial
coordinate $x$ by several units of $1\lambda /2$) and at a time $%
t_{localized}$ for which there is appreciable heating along the axial
direction but for which there is no flight (i.e., the atom remains localized
within the same axial well). The times $(t_{flight},t_{localized})$ are
indicated by the arrows in the top two panels of parts (a).

Perhaps the most striking aspect of the comparison of the spectral
distributions $\{{\cal N}_{flight}(\Omega ),{\cal N}_{localized}(\Omega )\}$
for the cases with and without flight is their remarkable similarity [in (b)
of Figs.\ 14 and 15]. Both display prominent peaks near $\Omega _{p}/
2\pi =f_{p}\simeq $500--600 kHz, which is in accord with the expected
frequency for large-amplitude oscillation in the axial potential, for
which the harmonic frequency $f_{0}^{(a)}\approx 430$ kHz (recall that 
frequency of atomic dynamics is half the frequency of the associated
variations in $\bar{n}(t)$). This result is also in accord with that from
Fig.\ 4(b) of Pinkse et al., for which their simulation leads to $1/
\tau _{p}\simeq $550 kHz for variations in the function $g^{(4)}$.

However, our analysis, as in the comparison of $\{{\cal N}_{flight}(\Omega ),%
{\cal N}_{localized}(\Omega )\}$ above, indicates that neither the
observation of a peak in ${\cal N}(\Omega )$ around $\Omega _{p}$ nor of
oscillatory structure in $g^{(4)}(\epsilon ,\tau ,\epsilon )$ around $\tau
_{p}\simeq 2\pi /\Omega _{p}$ is sufficient to justify direct
evidence for the atom moving along the cavity axis. \ Rather, peaks in $%
{\cal N}_{t_{i}}(\Omega )$ are ubiquitous around frequencies $\Omega
_{p}/2\pi \simeq $500--600, and appear whether the atom's motion is localized
(but heated) within a given axial well or whether the atom is in flight
across several wells. This feature follows from an analysis of the
full record of $\bar{n}(t)$ without the deleterious effects of finite escape
and detection efficiency, or of finite detection bandwidth. Such a result
suggests that the measurements of Fig. 4 in Ref\cite{rempe00a} are not in
and of themselves sufficient to establish unambiguous observation of atomic
motion across several wells of the cavity field standing wave.

\begin{figure}[h]
\centerline{\psfig{file=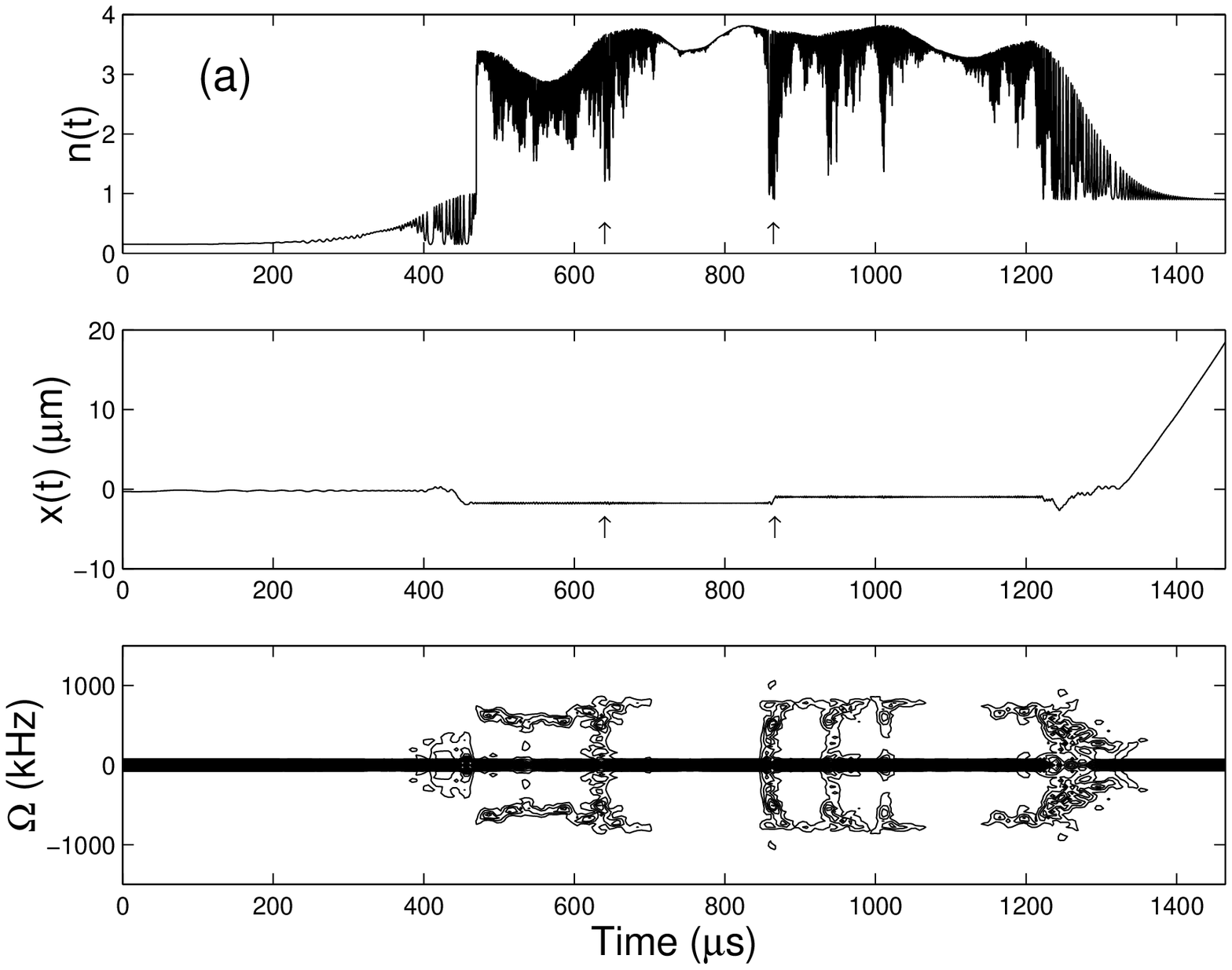,width=3.25in,height=2.54in}}
\centerline{\psfig{file=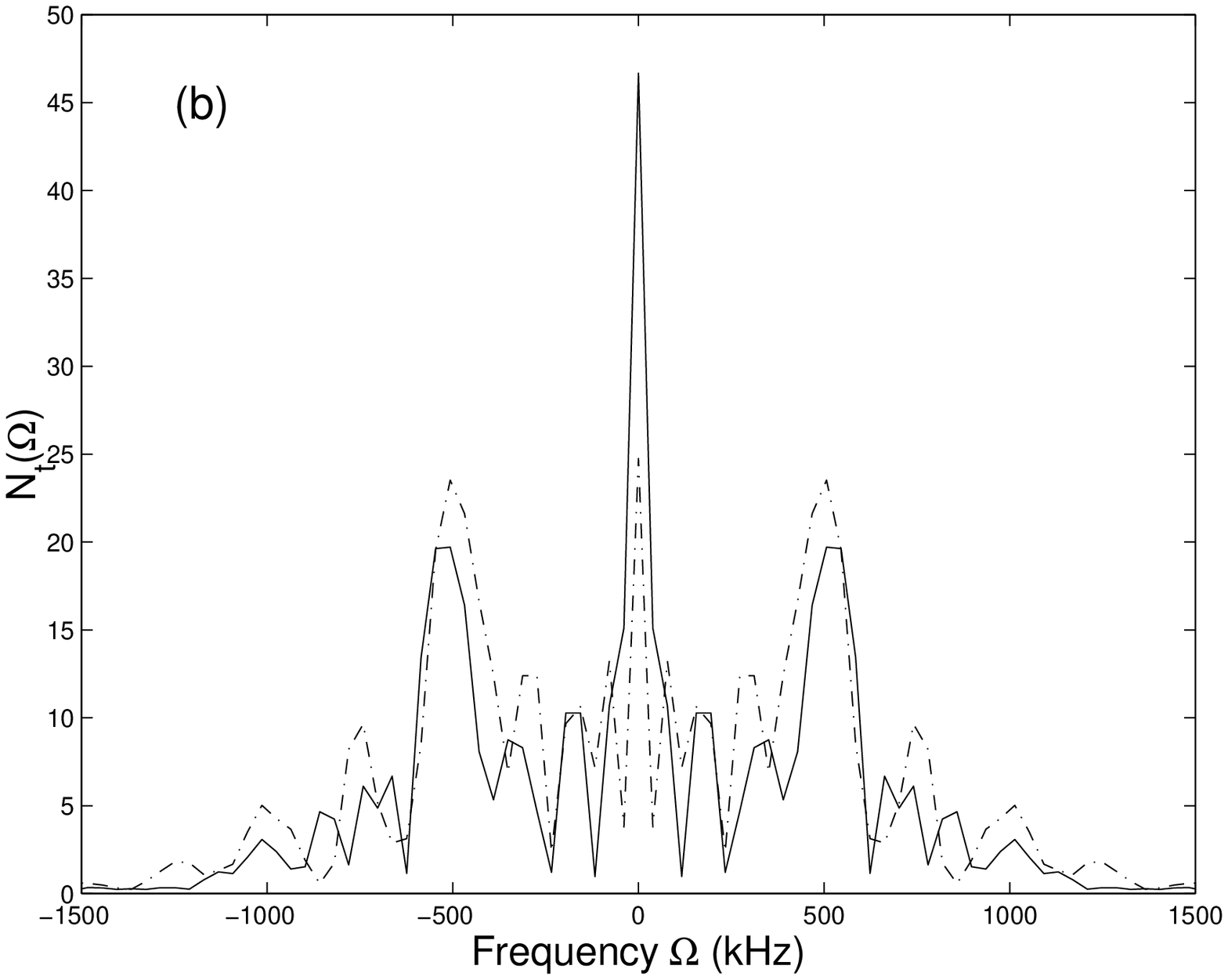,width=3.25in,height=2.54in}}
\caption{\narrowtext (a) Mean intracavity photon number $\bar{n}(t)$, axial
  position $x(t)$,
and a contour plot of the modulus of the windowed FFT ${\cal N}_{t}(\Omega )$
of $\bar{n}(t)$ for a simulated transit for the parameters of Pinkse
et al. (b) At the times indicated in (a), $|{\cal N}_{localized}(\Omega
)|$ is plotted corresponding to the arrow at $t_{localized}=652$ $\mu$s (solid
curve) and $|{\cal N}_{flight}(\Omega )|$ corresponding to the arrow at $%
t_{flight}=867$ $\mu$s (dash-dotted curve). There are apparently only minor
differences between these two spectra, which does not support the conclusion
about axial motion drawn from Fig.\ 4 in Ref. [8].}
\end{figure}

Our analysis does suggest that it may still be possible to distinguish
between axial motion confined within a well and flight along the cavity axis
through a more careful quantitative analysis of the respective spectral
distributions $\{{\cal N}_{flight}(\Omega ),{\cal N}_{localized}(\Omega )\}$%
. With reference to Figs.\ 14 and 15, note that a principal distinction
between these cases is that in the case of flight there is a large {\it %
decrease} of spectral content in the lowest frequency components around $%
\Omega =0$. This decrease reflects the fact that axial skipping causes
full-range variation in $g$, and thereby pulls down the time-averaged value
of transmission $\bar{n}(t)$. In addition, we note an {\it increase} in $%
{\cal N}_{flight}(\Omega )$ as compared to ${\cal N}_{localized}(\Omega )$
for Fourier components in a broad range around $\Omega _{p} / 2$ and
up to $\Omega _{p}.$ The increase appears to reflect atomic motion that,
during skipping, explores the full nonlinear (anharmonic) range of the axial
potential. These characteristics of the overall spectral distributions
seem to discriminate more reliably between flight and localized heating than
does a single-frequency peak criterion; they may still offer an avenue for
observing atomic skips across the standing wave.

\begin{figure}[h]
\centerline{\psfig{file=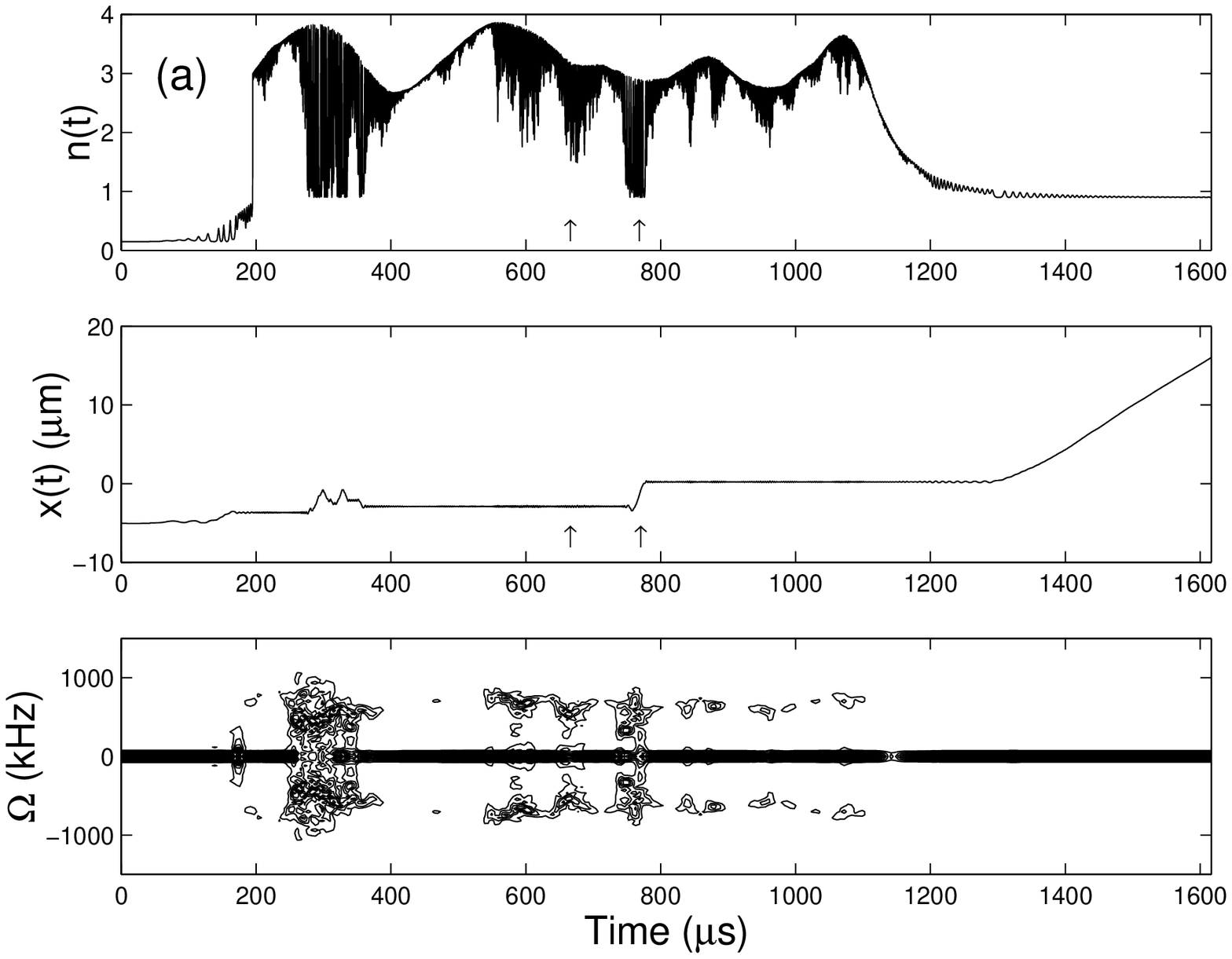,width=3.25in,height=2.54in}}
\centerline{\psfig{file=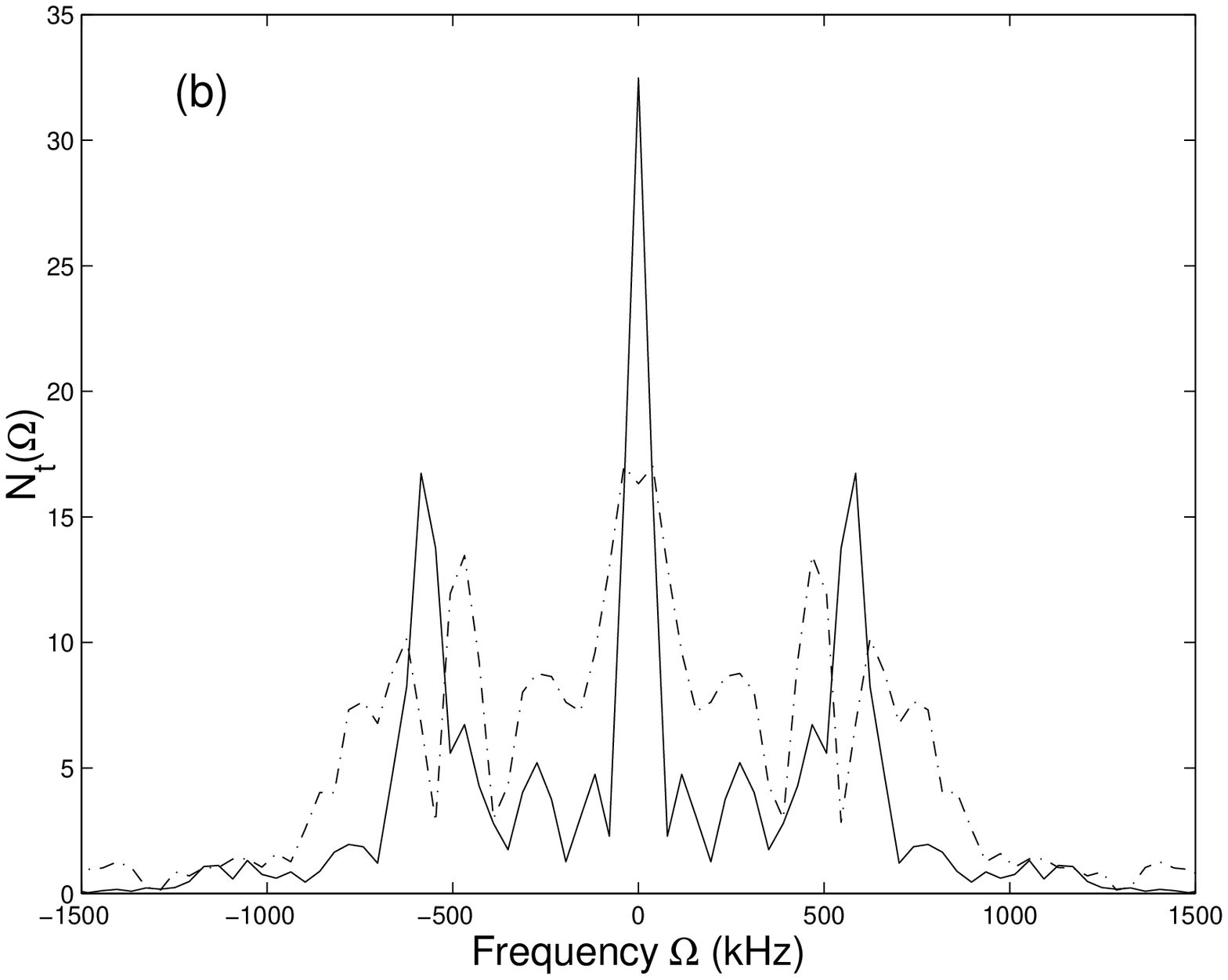,width=3.25in,height=2.54in}}
\caption{\narrowtext (a) Mean intracavity photon number $\bar{n}(t)$, axial
  position $x(t)$,
and a contour plot of the modulus of the windowed FFT ${\cal N}_{t}(\Omega )$
of $\bar{n}(t)$ for a simulated transit for the parameters of Pinkse
et al. (b) At the times indicated in (a), $|{\cal N}_{localized}(\Omega
)|$ is plotted corresponding to the arrow at $t_{localized}=673$ $\mu $s (solid
curve) and $|{\cal N}_{flight}(\Omega )|$ corresponding to the arrow at $%
t_{flight}=780$ $\mu $s (dash-dotted curve). See the text for discussion.}
\end{figure}

\section{Conclusions}

A principal objective of this paper has been to investigate the extent to
which light-induced forces in cavity QED\ are distinct from their free-space
counterparts. Our perspective has been to seek qualitatively new
manifestations of optical forces at the single-photon level within the
setting of cavity QED. Note that the importance of a quantum character for
the relevant fields or phenomena is not ensured by the statement that the
mean photon number $\bar{n}\sim 1$, since this is trivially the case in an
equivalent free-space volume for a field of the same intensity as that
inside the cavity.

As a starting point, we have presented comparisons between the effective
potential $U_{eff}(\rho ,x)$ in cavity QED and the corresponding free-space
potential, as well as of the diffusion coefficients in both contexts (Figs.\
3 and 4). Perhaps surprisingly, even in a regime of strong coupling as
in Ref.\cite{rempe00a}, there 
are only small differences between the cavity QED and free-space potentials
and diffusion coefficients. Note that the comparison of Fig.\ 4 includes
``the back action of the atom on the cavity field''\cite{rempe00a}, and yet
there are nonetheless no substantive differences between the cavity QED\ and
free-space cases for the experiment of Pinkse et al. Hence, although the
cavity QED interactions do bring a substantial advantage for atomic
detection within the cavity volume, we conclude that the claim of trapping
an atom with single photons in Ref.\cite{rempe00a} involves no new
characteristics unique to the cavity QED environment, with the conservative
forces and diffusion largely described by the well-known free-space theory
(Fig.\ 4). Friction which enhances trapping in this regime can be 
ascribed to cavity-mediated cooling effects\cite{horak1997a,vuletic2000a},
which are in 
themselves not uniquely features of the quantized-field treatment. However,
more analysis is required to determine if the observed effects of friction
do indeed rely on the cavity-field
quantization.

By contrast, for the experiment of Hood et al., a comparison of the
free-space theory and its cavity QED counterpart demonstrates that the usual
fluctuations associated with the dipole force along the standing wave are
suppressed by an order of magnitude. A semiclassical treatment of the cavity
field yields large diffusions like those calculated for the free-space trap.
\ Indeed, if it were not for the reduction of heating in the quantized
cavity QED\ case, an atom would be trapped for less than the period of a
single radial orbit before being heated out of the well for the parameters
of Ref. \cite{hood00}. \ Our calculations support the conclusion that the
suppression in dipole-force heating is based upon the Jaynes-Cummings ladder
of eigenstates for the atom-cavity system, which to our knowledge
represents qualitatively
new physics for optical forces at the single-photon level within the setting
of cavity QED.

In terms of a more complete analysis, the effective potential $U_{eff}(\rho
,x)$ and the diffusion coefficient $D(\rho ,x)$ are important ingredients in
the quasiclassical theory that we have developed for atomic motion in
cavity QED. By way of detailed, quantitative comparisons with the experiment
of Hood et al. in Sec.\ IV, we have validated the accuracy and utility of
our numerical simulations based upon the quasiclassical theory. As part of
this comparison, we have demonstrated agreement between experiment and
simulation for histograms of the duration of transit events, with mean $\bar{%
\tau}_{t}=340$ $\mu $s for the histogram in the triggered case of Fig.\ 9b
extended well beyond the mean $\bar{\tau}_{u}=92$ $\mu $s for the untriggered
case. Furthermore, $\bar{\tau}_{t}$ exceeds the transit time $\tau
_{0}=110$ $\mu $s\ for an atom to transit freely through the cavity mode. The
simulated trajectories of Fig.\ 6 together with the comparison of Fig.\ 10
for the experiment of Hood et al. strongly support the conclusion that
atomic motion is largely conservative in nature, with only smaller
contributions from fluctuating and velocity-dependent forces. Atomic motion
is predominantly in radial orbits transverse to the cavity axis. The
(suppressed) axial heating is important, but only towards the end of a given
trajectory leading to ejection from the trap. Knowledge of the time
dependence $\rho (t)$ for the radial coordinate (by way of the detected
field emerging from the cavity and the solution of the master equation) as
well of the confining potential $U(\rho ,0)$ allow an algorithm to be
implemented for inference of the actual atomic trajectory, as demonstrated
in Ref.\cite{hood00} and discussed in greater detail in Ref.\cite
{reconstruct}.

In the case of Ref.\cite{rempe00a}, numerical simulations for the parameters
appropriate to this experiment lead to histograms with mean $280$ $\mu $s in
the triggered case of Fig.\ 11(a) and $160$ $\mu $s for the untriggered case of
Fig.\ 11(b), which should be compared to the time $\tau _{0}=290$ $\mu
$s for an 
atom to transit freely through the cavity mode in this experiment. The
simulated transits of Fig.\ 7 indicate that atomic motion in this case is
dominated by diffusion-driven fluctuations in both the radial and axial
dimensions with friction playing an important role in the axial
direction. The character of the motion hampers inference of atomic
motion from the record of intracavity photon number. Axial heating leads to
repeated large bursts of axial excursions during an atomic transit, and hence
to large oscillations in the intracavity photon number $\bar{n}(t)$. The
envelopes of these oscillations have appreciable Fourier content in the
range of interest for observation of radial motion, so that there is not an
unambiguous signature for the radial motion in the record of $\bar{n}(t)$ on
short time scales, such as those presented in Ref.\cite{rempe00a}. Similarly,
the result by Pinkse et al. for hopping or flights over the antinodes of the
cavity standing wave is not substantiated by a closer inspection of the
Fourier content of the relevant signals. As documented in Figs.\ 14 and 15,
similar signals can be observed for an atom localized (but heated)\ within a
single standing-wave well. We emphasize that these conclusions concerning
the work of Ref.\cite{rempe00a} are based upon the analysis of several
hundred simulated trajectories, apparently well beyond the few cases
presented in that paper.

Beyond these comments directed to the prior work of Refs.\cite
{hood00,rempe00a}, we suggest that the capability for numerical simulation
of the quasiclassical model of atom motion in cavity QED should have
diverse applications. For example, we are currently applying the simulations
to the problem of feedback control of atomic motion. Given the capability to
infer an atomic trajectory in real time, it should be possible to apply
active feedback to cool the motion to the bottom of the effective potential $%
U_{eff}(\vec{r})$.

\section*{Acknowledgments}

We gratefully acknowledge the contributions of K. Birnbaum, J. Buck, H.
Mabuchi, S. Tan, and S. J. van Enk to the current research. This work was
supported by DARPA\ via the QUIC\ Institute administered by ARO, by the NSF,
and by the ONR.

\end{multicols}

\begin{references}
\bibitem{foothetvscount}  For the experiment of Ref. \cite{rempe00a}, the
free-space standing wave for comparison is assigned the same peak intensity $%
g_{0}\langle a^{\dagger }a\rangle (0,0)$ as the cavity field mode. \ The use
of peak $\langle a^{\dagger }a\rangle $ in this case vs peak $|\langle
a\rangle |^{2}$ in the case of Ref.\cite{hood00} reflects the fact that
photon counting was employed in Ref.\cite{rempe00a}, while detection was via
heterodyne in Ref.\cite{hood00}.
\end{references}
\end{document}